\documentclass[12pt,preprint]{aastex}

\newcommand{\beq}{\begin{equation}}
\newcommand{\enq}{\end{equation}}
\newcommand{\beqa}{\begin{eqnarray}}
\newcommand{\enqa}{\end{eqnarray}}
\newcommand{\cross}{\mathbf{\times}}

\shortauthors{Vestuto et al}
\shorttitle{Compressible MHD Turbulence}

\begin{document}

\title{Spectral Properties of Compressible Magnetohydrodynamic Turbulence from Numerical Simulations}

\author{Jason G. Vestuto$^{1}$, Eve C. Ostriker$^{1}$, and James M. Stone$^{1,2}$}
\affil{$^1$Department of Astronomy, University of Maryland \\
College Park, MD 20742-2421}
\affil{$^2$DAMTP, University of Cambridge \\ 
Silver Street, Cambridge, England CB3 9EW}
\email{vestuto@astro.umd.edu, ostriker@astro.umd.edu, jstone@astro.umd.edu}

\begin{abstract}
  
  We analyze the spectral properties of driven, supersonic
  compressible magnetohydrodynamic (MHD) turbulence obtained via
  high-resolution numerical experiments, for application to
  understanding the dynamics of giant molecular clouds. Via
  angle-averaged power spectra, we characterize the transfer of energy
  from the intermediate, driving scales down to smaller dissipative
  scales, and also present evidence for inverse cascades that achieve
  modal-equipartition levels on larger spatial scales. Investigating
  compressive versus shear modes separately, we evaluate their
  relative total power, and find that as the magnetic field strength
  decreases, (1) the shear fraction of the total kinetic power
  decreases, and (2) slopes of power-law fits over the inertial range
  steepen. To relate to previous work on incompressible MHD
  turbulence, we present qualitative and quantitative measures of the
  scale-dependent spectral anisotropy arising from the
  shear-Alfv\'{e}n cascade, and show how these vary with changing mean
  magnetic field strength. Finally, we propose a method for using anisotropy in velocity centroid maps as a diagnostic of the mean magnetic field strength in observed cloud cores.

\end{abstract}


\section{Introduction}

The molecular interstellar medium (ISM) is observed to be turbulent at
all spatial scales -- from the dense, small prestellar cores where the
turbulent Mach number ${\cal M}_s\equiv \sigma_v /c_s$ is order-unity,
to the low (mean) density, giant molecular cloud complexes (GMCs)
where ${\cal M}_s$ is 10 or more (e.g. \cite{Blitz1993}); here
$\sigma_v$ is the velocity dispersion and $c_s$ is the sound speed.
Such high Mach numbers imply a strongly compressible state on large
spatial scales, which is manifest in the inhomogeneous structure of
observed GMCs.  Although magnetic field strengths are observationally
difficult to obtain (see e.g. \cite{Crutcher99}), magnetizations at up to
kinetic-equipartition levels are possible for GMCs that form in a
flux-conservative way from the diffuse ISM.  GMCs may thus be thought
of as gaseous entities pervaded by compressible magnetohydrodynamic
(MHD) turbulence.  This turbulence is believed to have many important
effects on cloud structure and evolution, but remains only
incompletely understood.

In recent years, large-scale numerical MHD simulations have been
introduced as a way to explore the fundamental properties of
compressible MHD turbulence and hence interpret the dynamics of GMCs
(see e.g. \cite{Vazquez2000}).  An initial emphasis of these
studies was to assess the dissipation rate of compressible MHD
turbulence (e.g. \cite{Stone1998}, hereafter Paper II; \cite{MacLow1998}; 
\cite{Padoan1999}), with the general conclusion that magnetic 
fields do not appreciably lengthen the dissipation time compared 
to the ratio $L/\sigma_v$ associated in
an unmagnetized medium with either the crossing time to form shocks
(for compressible flow) or the turnover time of an eddy (for
incompressible flow).  Other studies have focused on characterizing
various aspects of the density, velocity, and magnetic field
structure, both intrinsic and as observable with spectral or continuum
diagnostics (see e.g. \cite{Ostriker2002} and references therein).

An important property of compressible MHD turbulence that has not
previously been extensively analyzed is the power spectrum -- namely,
how the fluctuations of dynamical variables (especially velocity and
the magnetic field) vary with spatial scale and angular direction.
\footnote{The work of \cite{Cho2002}, contemporaneous with our 
study, also analyzes some spectral aspects of compressible MHD
  turbulence.}  These spectral properties reflect how energy is
transferred from input scales to smaller (dissipation) scales, and
potentially also to larger scales.  Spatial variations in heating may
depend on the nature of the ``forward'' energy cascade (to small
scales), while the viability of tapping internal sources of mechanical
energy (such as stellar outflows) to provide large-scale turbulent
support of a cloud depends on the efficiency of the ``inverse'' energy
cascade.  Determining what signatures the input scale leaves on the
power spectrum may help identify the sources of turbulence. Because
spectral properties depend on the mean magnetization of the medium,
spectral characteristics inferred from molecular line data cubes or
from polarized extinction/emission maps may be useful for diagnosing
the magnetic field strength.  One of the most well-know
characteristics of turbulence in molecular clouds is the general
power-law increase of linewidth with spatial scale, as initially
described by \cite{Larson81}.  Analytical methods are currently being
developed (e.g. \cite{HS97}; \cite{Rosolowsky99};
\cite{OM2002}; \cite{LP2000}; \cite{Ostriker2001}) 
that may enable relatively
detailed measurements of the velocity power spectrum from molecular
line observations.

With the above motivations in mind, in this paper we analyze a
high-resolution (up to $512^3$) set of supersonic (${\cal M}_s=5$)
driven-turbulence, ideal MHD simulations to characterize their basic
spectral properties.  Using four different models with Alfv\'en speeds
in the mean magnetic field from zero up to $2 \sigma_v$, we focus in
particular on how the spectral properties depend on the mean
magnetization of the system.

The plan of this paper is as follows: We begin in \S 2 with a brief
review of theoretical models and numerical results on power spectra in
other regimes of turbulence (incompressible and/or unmagnetized), to
place the current work in context.  In \S 3 we describe the numerical
methods used to construct our models and the parameter sets we have
adopted.  Section 4 presents the results of our spectral
analyses. First, in \S 4.1, we examine the angle-averaged
(one-dimensional) power spectra of the magnetic and velocity fields,
for shear and compressive components separately and in combination.
Section 4.2 then analyzes the global, scale-dependent, anisotropy
introduced to the power spectral density (PSD) in the presence of a
dynamically-strong mean magnetic field.  Finally, in \S 5, we conclude
with a discussion of our results and comparison to the results from
other numerical simulations, theoretical model predictions, and 
current observational measures of turbulence.

\section{Spectral Properties of Turbulence}

Most commonly, ``turbulence'' refers to a complex state of randomly
fluctuating velocity fields.  Depending on the nature of the medium
and the strength of the turbulence, there may also be accompanying
fluctuations in the local magnetic field and in the density and temperature.  
The scale-dependent structure of the turbulence can be described in many
ways, with perhaps the simplest characterization in terms of power spectra --
i.e. the Fourier transforms of the fluctuating part of the fluid variables.  
In general, a power spectrum $P({\bf k})$ may depend on all three components 
of the wave vector $\bf k$.  If the turbulence is isotropic, then $P$ 
depends only on the magnitude $k\equiv (k_x^2 + k_y^2 + k_z^2)^{1/2}$ of 
$\bf k$ (also written as $k_r$), and a one-dimensional 
spectrum averaged over polar and azimuthal angles fully characterizes $P$.  
When there is a preferred direction in the medium, as for 
the case with a strong, large-scale magnetic field, then turbulence 
may be anisotropic.  For a single preferred direction $\hat x$, $P$ depends 
only on the components of $\bf k$ parallel and perpendicular to $\hat x$,
and may be characterized by a two-dimensional spectrum $P(k_\parallel,
k_\perp)$ averaged over cylinders centered on the $\hat x$ axis.

An equivalent way of characterizing the scale dependence of turbulence 
is in terms of the variance as a function of physical scale;  ``Larson's-Law''
empirical linewidth-size relations use this representation. 
For the case of isotropic turbulence with a power-law spectrum, 
the variance of the corresponding 
fluid variable will also increase as a power law as a function of the 
spatial scale over which it is averaged.  For example, if 
$v^2({\bf k})\propto k^{-n}$, then the velocity dispersion increases as a 
function of physical size $\ell$ as 
$\sigma_v(\ell) \propto \ell^{(n-3)/2}$. 
\cite{Larson81} originally reported a linewidth-size relation 
$\sigma_v(\ell) \propto \ell^{0.38}$; larger exponents $\sim 0.5-0.6$
have been reported for larger scales with more homogeneous data sets 
(e.g. \cite{Solomon87}; \cite{Brunt2002}). 
Because a number of issues remain unresolved in 
extracting the velocity power spectrum from molecular-line data cubes, at 
present we may regard observed turbulence in GMCs as broadly consistent
with a range of possible spectral indices.

The most familiar model of turbulence is that of \cite{K41},
conceived to describe the energy cascade from large to small (dissipation) 
scales in
incompressible, unmagnetized flows.  Local, isotropic interactions of
eddies with velocities $v({\ell})$ at physical scales $\ell$ have
durations $\ell/v(\ell)$.  An assumption of a conservative energy
cascade within an ``inertial range''
then implies $v^3(\ell)/\ell$ is scale-independent, so that
the velocity dispersion on the scale $\ell$ is 
$v(\ell)\propto \ell^{1/3}$.  
Assuming isotropy and converting to a power spectrum
using $\int_k d^3k' P({\bf k'}) \propto v^2(\ell)$ for $k=2\pi/\ell$,
the implied angle-averaged Kolmogorov power spectrum thus obeys the
scaling $P(k) \propto k^{-2/3} \times k^{-3} \propto k^{-11/3}$.  
Numerical simulations in the mildly compressive limit (decay models with 
${\cal M}<1$, and driven models with ${\cal M}\sim 1$) support the velocity 
scaling predicted in the Kolmogorov model while showing that the 
compressive component of energy amounts to 5-10\% of the total 
\citep{PPW94,PWP98,PPSW99}. Because the Kolmogorov model assumes 
an incompressible medium, it cannot be expected to apply at 
large scales in a molecular cloud, where the turbulence is
strongly supersonic.

At an opposite extreme from Kolmogorov's energy-conservative, incompressible 
turbulence cascade is the \cite{Burgers74} model
 of shock-dominated turbulence.  In 
this and related models, the presence of shocks transfers energy immediately
from large-scale motions to dissipative scales, rather than via a 
cascade through ``eddies'' at intermediate spatial scales.  The spectrum
$P(k)\propto k^{-4}$ associated with Burgers turbulence corresponds to 
the Fourier transform of a collection of step functions representing
many individual shock profiles. Conversion from $k$-space to the spatial domain yields $v(\ell)\propto \ell^{1/2}$. 


Neither the Kolmogorov nor the Burgers model incorporates magnetic
fields, which may be important to the mode of energy transfer through
scales -- particularly if the total magnetic energy density nears
that in the turbulent velocity field.  Approaches to theoretical
modeling of magnetized turbulence have focused on the incompressible
case, in the limit in which fluctuations in the magnetic field are small
compared to its mean value; an excellent current review is contained in \cite{Cho2002b}.  
The physical picture in the early 
theory of \cite{IK63} and \cite{IK65} (collectively, ``IK'') 
is based on energy transfer to smaller
scales initiated by collisions of counter-propagating Alfv\'en wave packets.  
An assumption of an energy-conservative cascade and isotropy
(with the implication that many collisions are needed to dissolve a 
wavepacket) yields a spectrum $P(k)\propto k^{-7/2}$.  However, consideration
of the resonant conditions for three-wave interactions among Alfv\'en 
waves suggests that the cascade will in fact {\it not} 
be isotropic, but instead should
preferentially transfer energy in the direction perpendicular to the 
mean magnetic field \citep{SMM83} (hereafter SMM83).
Numerical simulations in 2D and 3D by SMM83 and \cite{OPM94} directly 
demonstrated that spectral anisotropy extended in the $k_\perp$ direction 
is indeed present; this anisotropy increases toward smaller spatial scales.

A model of incompressible MHD turbulence that uses the
colliding-wavepacket picture of IK, while replacing the unphysical
assumption of isotropy with the primarily perpendicular cascade of SMM83, was
proposed by Goldreich \& Sridhar (1995, GS; see also \cite{GS97}).  
GS introduce the concept of a ``critically-balanced''
cascade in which the nonlinear (``eddy turnover'') time of a
wavepacket and its propagation time remain comparable, so that
$k_\parallel v_A \sim k_\perp v(k_\perp)$.  Combining this relation
with an energy-conservative near-perpendicular cascade ($k_\perp
v^3(k_\perp) \sim constant$) yields a scale-dependent anisotropy law
$k_\parallel \propto k_\perp^{2/3}$.  Numerical simulations in the 
incompressible limit by \cite{Cho2000}, \cite{MG2001}, and \cite{Cho2002a} 
have verified (via structure functions for coordinates aligned 
with the local $\bf B$-field) that the expected scale-dependent anisotropy is
present.  The power spectrum from these incompressible 
MHD simulations is similar to or slightly steeper than the Kolmogorov 
value (see also \cite{Muller2000}).

To our knowledge, no analytic model of comparable conceptual
simplicity to those of Kolmogorov, Burgers, or Goldreich \& Sridhar in
their respective regimes presently exists for compressible MHD
turbulence.  In practice, elements of the latter two theories may both
apply: at large scales, velocities are strong enough to drive shocks,
while at small scales (away from shocks), velocity perturbations are
small compared to the sound speed and magnetic perturbations are small
compared to the mean magnetic field.  However, the overall result
cannot be a simple hybrid for the system as a whole, because there is
no clean separation of scales when shocks produced by large scale
motions transfer energy immediately to the dissipation scale.  Results
from lower-resolution numerical work indicate relatively steep power
spectra \citep{MacLow2000}, while \cite{Cho2002} have
recently reported results showing anisotropy of the expected sense
(see also earlier work in the weakly compressible limit by 
\cite{MGOR96,MOGH98,OMG98}).
The analysis of the present paper focuses on comparing turbulent
spectra in strongly compressible 
models (${\cal M}_s=5$) with varying background magnetic field strength,
based on simulations with the highest resolutions ($512^3$) performed
to date.

\section{Models and Parameters}

Our turbulent cloud models are obtained on a spatial grid of $256^3$
zones (models A, B, C, and D) or $512^3$ zones (models A2, C2). These
models were studied in previous work (\cite{Stone1998}, hereafter Paper II),
which focused on the steady-state dissipation rate of turbulence.
Models A-D here represent the same simulations with dissipation
characteristics listed in Table 1 of Paper II, while models A2 and C2
are higher resolution counterparts.

Each model begins as a cubic box of dimension L, initially filled with
a stationary, isothermal plasma of uniform density
$\rho_{0}=\bar{\rho}$, and threaded by a uniform
magnetic field $\bf{B}$ = $B_{0}\hat{x}$. For
each model, the initial magnetic field strength $B_{0}$ is
parameterized by $\beta \equiv
c_{s}^{2}/v_{A,0}^{2}=c_{s}^{2}/(B_{0}^{2}/4\pi\bar{\rho})$, being
proportional to the ratio of the thermal to magnetic pressure. We
consider models representing a strong field case ($\beta =0.01$;
models A, A2), a moderate field case ($\beta =0.1$; model B), a weak
field case ($\beta =1$; models C, C2), and a purely hydrodynamic case
($\beta = \infty$, model D).

We evolve each model in time using the ZEUS code (see \cite{Stone1992a}, 
\cite{Stone1992b}) to integrate the compressible, ideal MHD equations
\beq
\frac{\partial \rho}{\partial t} = -\nabla\cdot(\rho \bf{v}),
\enq
\beq
\frac{\partial {\bf v}}{\partial t} + {\bf v}\cdot{\bf \nabla}{\bf v} = 
- \frac{{\bf \nabla} P}{\rho}  - \frac{{\bf \nabla}{B^{2}}}{8\pi\rho} 
+ \frac{{\bf B}\cdot{\bf\nabla}{\bf B}}{4\pi\rho},
\enq
\beq \frac{\partial {\bf B}}{\partial t} = {\bf \nabla}\cross({\bf
  v}\cross{\bf B}).  
\enq 

We adopt an isothermal equation of state $P = c_{s}^{2}{\rho}$,
where the sound speed $c_{s}$ is constant in space and time. We view
this as an acceptable approximation for modeling most of the 
gas within a GMC (see discussion in \cite{Ostriker1999}). 
We do not include any explicit viscous or resistive terms 
other than shock-capturing artificial viscosity in the MHD equations, 
and thus dissipation is purely numerical.

Boundary conditions for all of our models are periodic. For a given
model, $\beta$ remains constant in time due to the periodic boundary
conditions, which maintain volume averaged
$\langle\delta\mathbf{B}\rangle$=0. The initial uniformity of $B_{0}$
and $\rho_{0}$ also implies a spatially and temporally constant
mass-to-flux ratio, modulo the numerical reconnection effect that may occur if magnetic flux elements of opposite signs are advected into a single cell.

The initial velocity field of each model is $\mathbf{v_{0}} = 0$, with
turbulent driving implemented by introducing velocity perturbations
$\delta\bf{v}$ at time intervals $\Delta t = 0.001 t_{s}$ (where
$t_{s}=L/c_{s}$ is the sound crossing time over the box).  The
perturbations follow a Gaussian random distribution with Fourier power
spectrum 
\beq 
\langle |\delta \mathbf{v(k)}|^{2} \rangle \propto
k^{6}\exp{\left(-8 k/k_{peak}\right)}.  
\enq 
Peak driving occurs at
$k_{peak}L/2\pi=8$ ($\lambda_{peak}=L/8$) for all models except 
model $C_{k4}$, for which $k_{peak}L/2\pi=4$ so that it may later 
be used in a rescaling comparison with model C2. 
For all of our models, the energy input rate
$\dot{E}$ due to driving, given by $\dot{E}/\rho_{0}L^{2}c_{s}^{3}=1000$, 
is constant. The driving is incompressive, with the
constraint imposed that $\bf{\nabla}\cdot\delta\bf{v}$=0. The
perturbations are normalized such that no net momentum is added. 
This is accomplished by adding a spatially-uniform 
component of velocity to compensate for any net momentum associated 
with the sum of sinusoidal velocity increments,  $\int\rho\delta\bf{v}$=0 . 
The resulting saturated-state energies associated with each model 
were presented in Paper II and will be further analyzed later in \S 3.

Additional model parameters are provided in Table 1 (see also Paper
II), including the turbulent Mach number ${\cal M}_{s}\equiv\sigma_{v}/c_{s}$ 
and the Alfv\'{e}n Mach numbers ${\cal M}_{A,0}\equiv
\sigma_{v}/v_{A,0}$, ${\cal M}_{A,rms}\equiv \sigma_{v}/v_{A,rms}$.  
Here the Alfv\'{e}n speed
$v_{A,0}^{2}\equiv B_{0}^{2}/4\pi\bar{\rho}$ associated with the 
mean magnetic field is distinguished from
$v_{A,rms}^{2}\equiv\langle B^{2}\rangle/4\pi\bar{\rho}$, calculated
via a mass-weighted average of the squared Alfv\'en speed in each cell, 
$|{\mathbf B}_{0} + \delta{\mathbf B}|^2/4\pi\rho$, over the box. The 
velocity dispersion is calculated by a mass-weighted average over the box,
$\sigma_{v}^{2}\equiv \langle|{\bf v}|^{2} \rho/\bar{\rho}\rangle$. The sonic Mach numbers
are typically ${\cal M}_{s} \sim 5$, while values for ${\cal M}_{A,0}$ vary
from $\sim 0.5$ for the strong field models up to 5 for the weak
field models. The total turbulent energies listed in Table 1 are defined as $E_{K}=(1/2)\int d^{3}x |{\mathbf v}|^{2}\rho$ and $\delta E_{B}=(1/8\pi)\int d^{3}x (|{\mathbf B}|^{2}-B_{0}^{2})$, respectively. Dynamical time scales 
of interest include the flow crossing time $t_{f}(L)=L/\sigma_{v}$ and 
the Alfv\'{e}n wave crossing time $t_{A}(L)=L/v_{A,0}$ over the box, 
which are given by $t_{s}/{\cal M}_{s}$ and $t_{s}\beta^{1/2}$, respectively. 
Note that in Paper II, we frequently made use of the flow crossing time 
at the driving scale, $t_{f}(\lambda_{peak})=t_{f}(L)/8=t_{s}/8{\cal M}_{s}$. 
The values reported in Table 1 are for the saturated state 
at late times when the driving and dissipation have reached a balance. 
\footnote{Values for the $256^3$ models are averaged over times $t=0.25t_s$ and $0.30t_s$. Due to computational expense, the $512^3$ models were not evolved for as many time steps. Values for model A2 and C2 are given at times $t=0.10t_s$ and $t=0.15t_s$ respectively. These times are well beyond the time $t_{f}(\lambda_{peak})\approx 0.025t_{s}$ needed to allow for the development of shocks arising from bulk flow motions.}

For the purpose of relating the dimensionless units of our simulations
to physical units, we give the sound crossing time as
$t_{s}=53(L/10{\rm pc})(T/10{\rm K})^{-1/2}{\rm Myr}$, and the
Alfv\'{e}n wave crossing time as $t_{A}=7.6(L/10{\rm
  pc})(n_{H_{2}}/10^{2}{\rm cm}^{-3})^{1/2}(B_{0}/10{\rm \mu
  G})^{-1}{\rm Myr}$. The value of $\beta$ for a given model
determines the corresponding physical value of the initial magnetic
field strength as $B_{0} = 1.4 {\rm \mu G} \beta^{-1/2}(T/10{\rm
  K})^{1/2}(n_{H_{2}}/10^{2}{\rm cm}^{-3})^{1/2}$, while the sound
speed is $c_{s}=0.19{\rm km s^{-1}}(T/10{\rm K})^{1/2}$. 
The driving power input per volume is then given by $\dot{E}/L^3=1.04\times 10^{-25}{\rm ergs\ cm^{-3} s^{-1}}(n_{H_{2}}/10^{2}{\rm cm}^{-3})(L/10{\rm pc})^{-1}(c_{s}/0.2{\rm km s^{-1}})^{3}$. 
Because the current simulations do not incorporate self-gravity, 
the length scale $L$ is arbitrary.

\section{Results}

In this section, we present the results of our spectral analyses. \S
4.1 focuses on angle-averaged spectral properties; we evaluate
inertial-range slopes for various components of the power, and assess
how these depend on $B_{0}$. In \S 4.2, we turn to the issue of
characterizing anisotropy in the power spectra. Spectral analysis was
performed on the models using a Danielson-Lanczos fast Fourier
transform (FFT) and computing the one-sided power spectral density
(PSD).

A question of some interest is the relative overall importance and
scaling behavior of the compressive versus the shear modes of
supersonic MHD turbulence. To compare these, we define separately the
shear and the compressive component of the velocity PSD. These are
computed in $k$-space, after the FFT has been performed, as

\beq
v_{shear}^{2}(\bf{\vec{k}}) = |\bf{\hat{k}}\cross\bf{v}(\bf{\vec{k}})|^2
\enq
\beq
v_{comp}^{2}(\bf{\vec{k}}) = |\bf{\hat{k}}\cdot\bf{v}(\bf{\vec{k}})|^2
\enq

We investigate the behavior of these separate components in 
both \S 4.1 and 4.2. In Table 1, we give the volume-integrated energies associated with these two components, for all models.

\subsection{Angle-averaged power spectra}

To study power spectra as a function of length scale only (i.e.
angle-averaged), we bin in spherical shells in $\mathbf{k}$-space so
as to generate PSD's as a function of $k=k_{r} \equiv (k_{x}^{2} + k_{y}^{2}
+ k_{z}^{2})^{1/2}$. The resulting functions are the shell-averaged
(average spectral energy per mode) power spectrum $P(k)$ and can be
related to the shell-integrated (total in a spherical shell)
energy spectrum as $dE(k)=4\pi k^{2} dk P(k)/2$.  

For model comparison, we will describe each of the resulting
$P(k)$ curves qualitatively in terms of four distinct ranges.
These are, in order of increasing $k$ (decreasing length): the
environmental range, driving range, inertial range, and dissipative
range. The environmental range extends from the largest spatial scales
(the computational box) to the scales at which the forcing is applied.
For saturated, driven turbulence, the $P(k)$ curve is nearly flat
through these largest scales -- i.e., there is near energy equipartition
among modes. As $k$ increases, there is a turnover
shortly after $k > k_{peak}$, where $k_{peak}$ defines the maximum in
the forcing function (eq. 4), indicating the start of the inertial
range. In the inertial range of $k$, the curve $P(k)$ is often fit well by
a power-law, with slopes varying roughly from -3.3 to -4.9, depending
on model parameters. For the present simulations, the inertial range typically spans from $kL/2\pi$ = 10 to 30 or 40, but is extended to 60 for the 
higher resolution models. In some cases, the choice of inertial range 
is somewhat subjective; the details of such cases are included in the 
discussion that follows. Finally, as $k$ increases still, the curve turns steeper once again at what we identify as the (purely numerical) dissipation range. 
Through the dissipation range, slopes tend to range between -6.5 and -7. 
The dissipation range is determined by purely numerical effects, as the code does not explicitly include dissipation terms except for shock-capturing artificial viscosity in the MHD equations. 
This generic form for $P(k)$ applies to both the kinetic and
magnetic components of $P(k)$, for both strongly and weakly
magnetized models, as seen e.g. in Figures 1a,b and Figures 2a,b,
respectively.

The primary quantitative measure we evaluate from the $P(k)$ in
our models is the slope in the inertial range. Table 2 shows the
exponent $n$ of power-law fits, $P(k) \propto k^{-n}$, in the
inertial range for the PSD of 
the (specific) kinetic energy $P_{\rm K}(k)\equiv v^{2}(k)/2$, 
the magnetic field energy $P_{B}(k)\equiv \delta B^{2}(k)/8\pi\bar{\rho}$,
the combination $P_{turb}(k)\equiv P_{\rm K}(k)+P_{B}(k)$, 
the incompressive component of (specific) kinetic energy 
$P_{shear}(k)\equiv v_{shear}^{2}(k)/2$, 
and the compressive component of (specific) kinetic 
energy $P_{comp}(k)\equiv v_{comp}^{2}(k)/2$.
The values presented in Table 2 for the $256^3$ models represent 
the analysis of an averaged data set obtained from two time snap-shots,
t=0.25$t_{s}$ and 0.30$t_{s}$ (corresponding to 1.25$t_{f}(L)$ and
1.5$t_{f}(L)$ or 10$t_{f}(\lambda_{peak})$ and
12$t_{f}(\lambda_{peak})$). The values reported for the $512^3$ models A2 and C2 are at times $t=0.10t_s$ and $t=0.15t_s$, respectively. 

While many spectra from the models exhibit an easily identifiable
power-law range, giving typical errors in slope fits $\pm 0.1$, some
important exceptions warrant detailed discussion.  The PSD curves for
density $\rho$ (not shown) have no discernible inertial range for any
of the models. Curves of $P_{B}(k)$ typically have very little
inertial range, and thus there is some ambiguity in choosing the
boundaries of the range to fit; typically a range within $kL/2\pi$=10
to 30 was used. The values cited were generated by averaging several
fits around the center of the most likely power-law, resulting in
slope fit errors of $\pm 0.2$. In several cases for $P_{\rm K}(k)$ and
$P_{shear}(k)$, particularly those of the weak field model, there is a
slight ``break'' in the middle of what we have identified as the
inertial range, with shallower slope at larger $k$. In these cases,
two approaches were taken to fitting the inertial range: (1) an single
average over the entire range and (2) fitting the two subranges on
either side of the break separately. Results of the first method are
cited in Table 2. Details of the second method are given in the
discussion of the general results which follows.

One striking effect is that the slopes characterizing the inertial
ranges of $P_{turb}(k)$, $P_{\rm K}(k)$, and $P_{shear}(k)$ all appear 
to scale with $\beta$, while those of the magnetic field and compressive 
velocity do not exhibit trends which are as definitive.  
Values for the PSD slopes of $P_{\rm K}(k)$ steepen 
as -4.0, -4.3, -4.7, -4.8 for $256^{3}$ models of
increasing $\beta$ = 0.01, 0.1, 1.0, and $\infty$, respectively. 
Steepening of the same sense is also evident for models A2, C2, with $P_{\rm K}(k)$ slopes -3.8 and -4.3, respectively.
The presence of a magnetic field thus appears to reduce intermediate-scale
dissipation of energy, in that the inertial range slopes
become flatter as one goes from models of no or small magnetic field
to those with greater mean field strength.

Spectra $P_{B}(k)$ and $P_{comp}(k)$ are notably different from kinetic energy spectra. The slopes of $P_{B}(k)$ vary only between -3.6 and -3.3 as one moves from the strong (A2) to weak (C2) field high-resolution models 
(with similar behavior at lower resolution). 
As the mean field weakens, the inertial range of $P_{B}$ becomes 
progressively harder to identify, being almost entirely unrecognizable 
for the weak field models. 
The inertial range of $P_{comp}(k)$ is much broader 
than that of other variables, particularly for the strong field model 
where the power-law fits extend over nearly twice the range 
of the corresponding fits for $P_{shear}(k)$. However, for moderate 
and weak-field low resolution models, the extent of the inertial range of $P_{comp}(k)$ is reduced again such that an average of fits is needed 
and the resulting errors in the indices reported are $\pm 0.2$.

Notice that there are two main differences between the spectrum of low
resolution models and those for their high resolution counterparts.
First, the indices in Table 1 for low resolution models tend to be
roughly 0.2 steeper than those of the high resolution models. In some
cases, this is due to the difference in identification of the inertial
ranges. For the low resolution model, the inertial ranges are of
lesser extent, which tends to result in fit being placed either closer
to the turnover from the driving range or the turnover into the
dissipation range; both cases lead to an artificial steepening of the
fit. Second, several of the curves for $P_{shear}(k)$ and $P_{\rm
  K}$(k), for both high and low resolution models, exhibit a slight
``break'' in the inertial range, as previously remarked. The
consequence is that the average fit made through this break results in
an apparent discrepancy between the indices reported for high and low
resolution counterparts.

The results of fitting either side of the slight spectral break within
the inertial range with separate power-laws reveals that in fact there
is agreement between the high and low resolution models, with some
interesting caveats.  The resulting indices of $P_{\rm K}(k)$ are
4.0,4.3,(4.9:4.5), and 4.8 for models A, B, C, and D, respectively,
with the notation $(n_{l}:n_{r})$ giving the indices of fits to the
left and right of the break (when one exists) with individual errors
of $\pm 0.1$. Compare these to the indices found for high resolution
counterparts A2 (4.0:3.5) and C2 (4.9:3.9), again with individual
errors of $\pm 0.1$. The fits to the left of the break within the
inertial range are in agreement.  The case of $P_{shear}(k)$ is
similar. Average and ``left-right'' indices $n$ and $(n_{l}:n_{r})$ of
fits to $P_{shear}(k)$ for low resolution models are 4.0, (4.5:4.0),
(5.0:4.5), 4.9 for models A, B, C, and D, respectively. Again,
comparing with the indices for the high resolution counterparts A2
(3.8:3.5) and C2 (5.0:4.0) reveals agreement to the left (smaller $k$)
of the break within the inertial range. For $P_{B}(k)$, the two-part
slopes are (3.5:3.7) for model A2, which may be compared with 3.6 for
the ``average'' slope, and 3.5 for model A.

Note that when a break is apparent within the inertial range, it occurs
well beyond the turnover that separates the driving range from the
inertial range. Also, the portion of the spectra to the left of the
break do not in any way mimic the input driving spectra. Although the
current simulations do not have sufficient resolution to be definitive
on this point, we speculate that the spectra to the right of the
break (larger $k$) represent the true asymptotic regime, 
while the spectra to the left of the break (smaller $k$) 
represent a transition between driving and asymptotic regimes. 
The steeper slope in the transition range may reflect 
the envelope of wave distributions in the process of
relaxing, via interactions, from the input spectrum to the asymptotic spectrum. This relaxation involves spreading in $k$-space accompanied by a decline in
the peak amplitude.

\begin{figure}
\center
\epsscale{1.}
\plotone{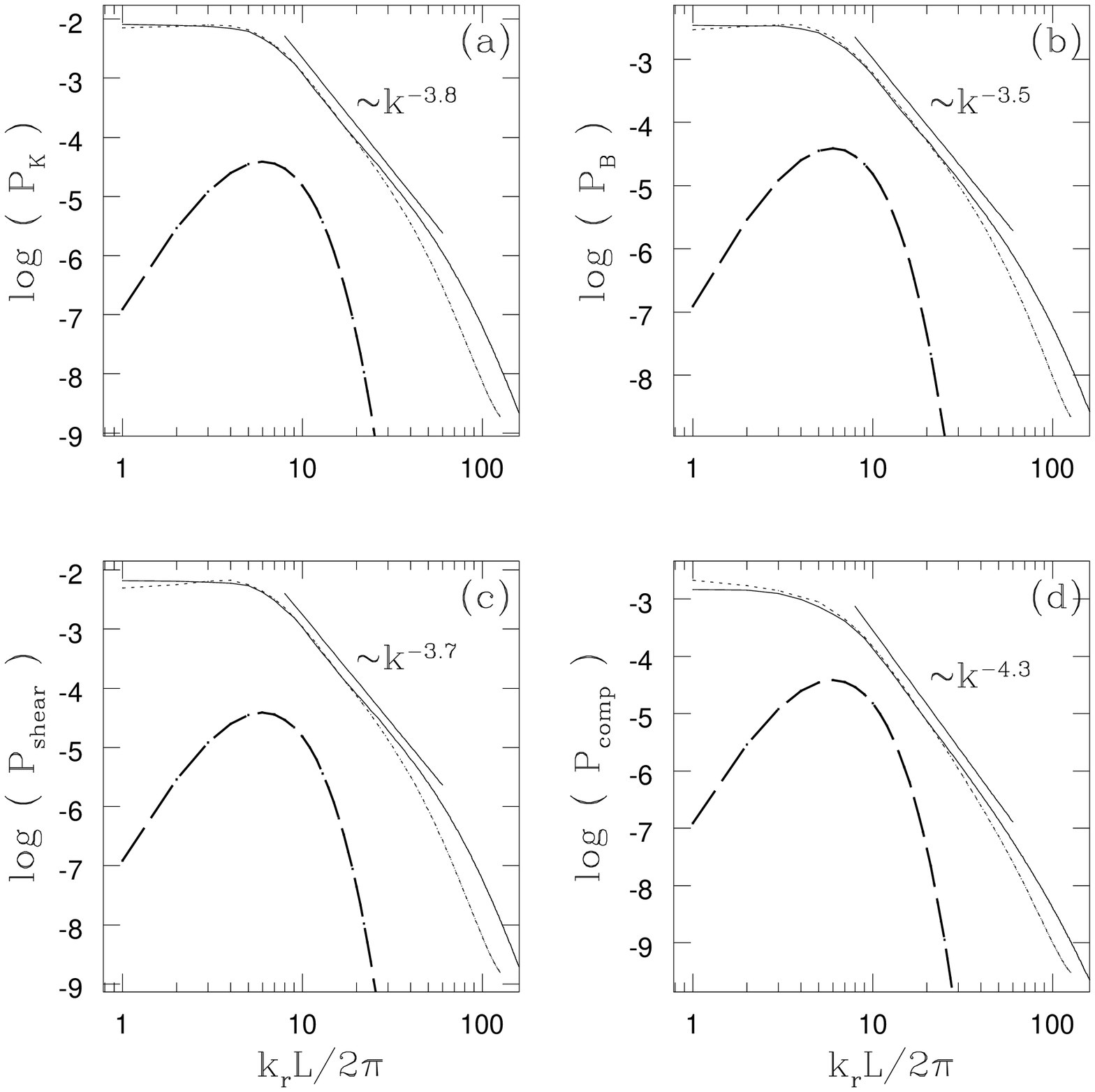}
\caption{Power spectral density in strong field case ($\beta=0.01$). 
  Spherically binned $P(k_r)$ of (a)
  $P_{\rm K}(k)$ , (b) $P_{B}(k)$, (c) $P_{shear}(k)$, and (d) $P_{comp}(k)$. Both
  $256^{3}$ (dotted curve) and $512^{3}$ (solid curve) resolution
  models are shown, with power-law fits (solid line) and indices (as
  labeled) for the $512^{3}$ model in the inertial range. Heavy dashed line shows the input
  spectral driving shape (eq. 4) with arbitrary amplitude.
\label{fig:plot1}}
\end{figure}

We now consider two examples to examine more specific effects which 
the magnetic field strength has on the form of the PSD curves. Figure
\ref{fig:plot1} illustrates the strong magnetic field case, with the
spherically binned PSD for a high-resolution $512^3$ model overlayed
with that of the lower resolution $256^3$ model (models A2 and A
respectively, both at time $t=0.10t_s$), both of $\beta$=0.01. 

In Figure \ref{fig:plot1}, there is clear indication of an inertial
range between $kL/2\pi=10$ and 60 for model A2, showing the cascade of
power from larger scale to smaller, dissipative scale at $kL/2\pi
>60$.  The power-law fits to the inertial ranges of $P_{\rm K}(k)$,
$P_{shear}(k)$, and $P_{B}(k)$ indicate scalings consistent with that
of Kolmogorov turbulence ($n=11/3$) or shallower, while that of
$P_{comp}(k)$ is significantly steeper.  Notice from Figure 1 that the
lower resolution model (A) has a more limited inertial range than that
of the high-resolution model.  The doubling of the resolution in
effect doubles the extent of the inertial range, because of the
greater separation between the driving scales and the dissipative
scales associated with the computational grid.

\begin{figure}
  \center \epsscale{1.}
\plotone{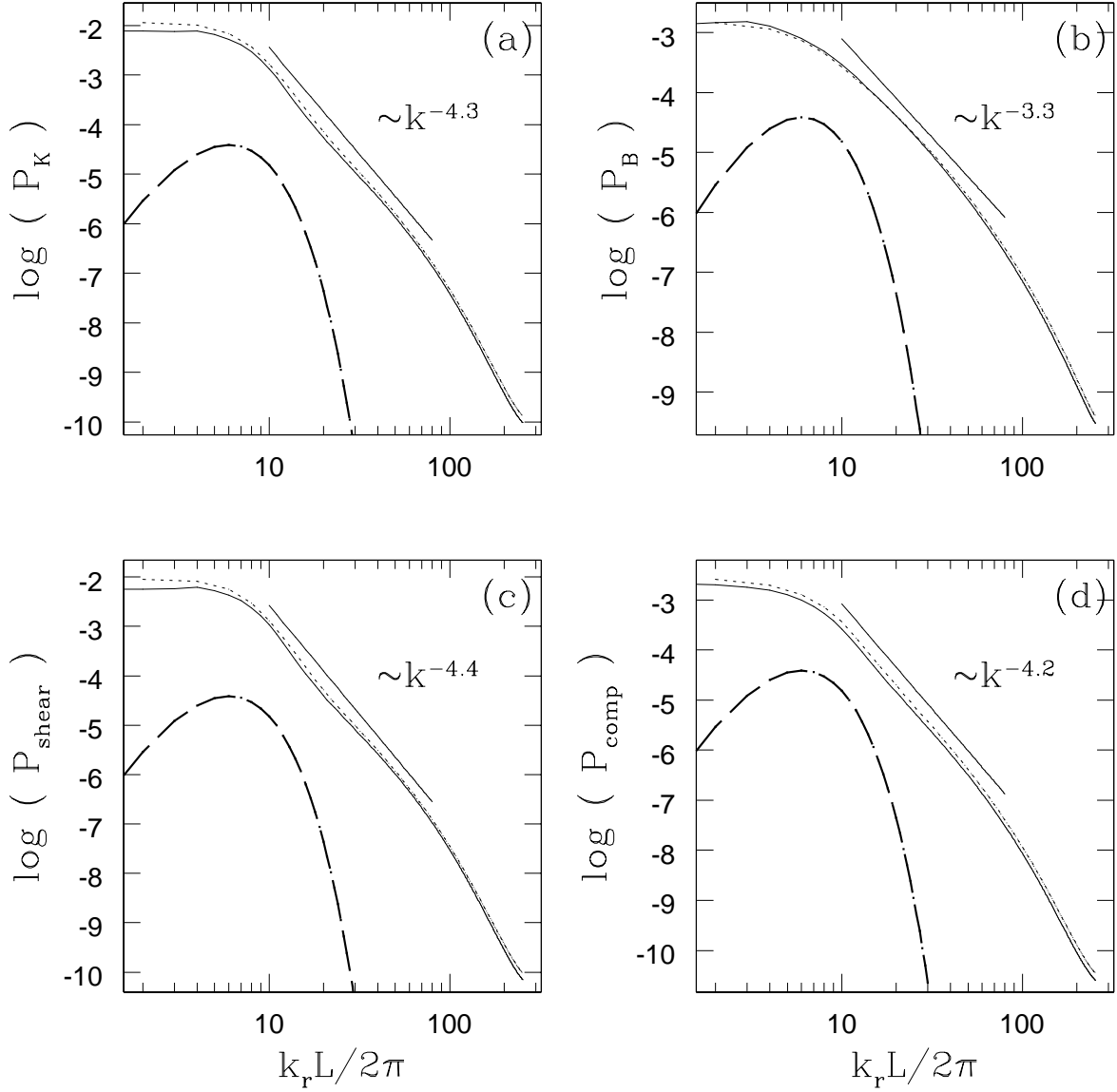}
\caption{ $P(k_r)$ in weak field case ($\beta=1.0$). Same quantities and 
  fits as in Figure 1.
\label{fig:plot2}}
\end{figure}

This interpretation is supported by Figure \ref{fig:plot2}, which
illustrates the weak field model. Here we again plot $P(k)$ curves
for a high-resolution model, with $k_{peak}L/2\pi=8$. We overlay these
with the spectrum from a $256^3$ model which was driven with
$k_{peak}L/2\pi=4$, rather than 8 as before. Here, both models are shown at $t=0.15t_s$. For the $256^3$ model, we
have rescaled $k$ by a factor of 2, and rescaled the magnitude by a
factor of 10, in order to compensate for the change in $k_{peak}$ and
the difference in total energies of each model, respectively. Note
that the inertial ranges are of the same extent and that the curves
practically line up, because the ratio of peak driving scale to
dissipation scale are identical.

\begin{figure}
  \center \epsscale{1.}
\plotone{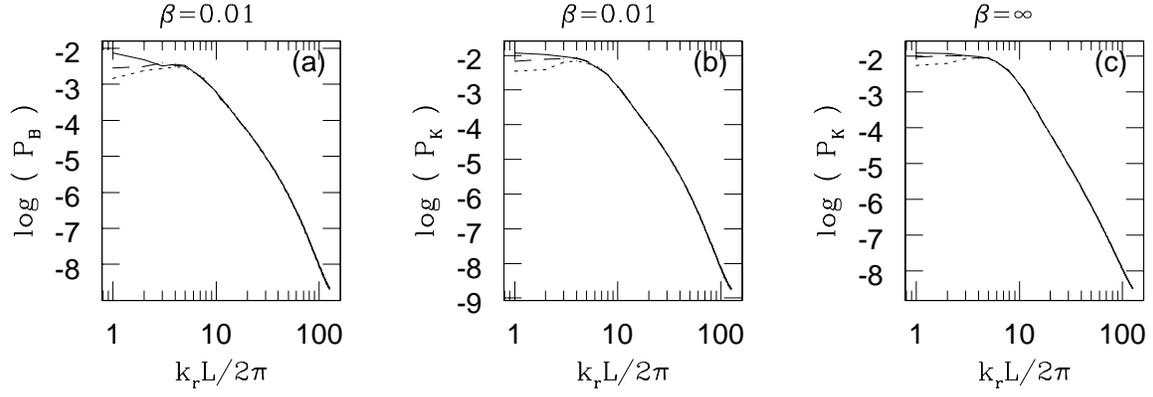}
\caption{Time series of $P(k_{r})$ for 
  (a) $P_{B}(k)$ and (b) $P_{\rm K}(k)$ for the strong-field case ($\beta=0.01$,
  model A) and (c) $P_{\rm K}(k)$ for the unmagnetized case
  ($\beta=\infty$, model D), at times equal to $0.05t_s$ (dotted
  curve), $0.1t_s$ (dashed curve), $0.3t_s$ (solid curve). 
\label{fig:plot4}}
\end{figure}

While driving at lower wave numbers evidently provides a larger
inertial range on the $k > k_{peak}$ side, our standard choice of
$k_{peak}L/2\pi=8$ significantly larger than unity allows us to
investigate inverse-cascade effects, that is, transfer of energy from
the driving scale to the larger ``environmental'' scales.  Evidence of
the growth of power on large scales is presented in Figure
\ref{fig:plot4}, which shows time series of magnetic and kinetic power
spectra from successive model snap-shots (times $t=0.05t_s$,
$0.10t_s$, and $0.30t_s$). For both the strong-field and unmagnetized
models, power at low $k$ increases over time. The unmagnetized and
magnetized models have similar, nearly flat, $P_{\rm K}(k)$ at low $k$,
whereas $P_{B}(k)$ rises slightly toward $k=0$ for the magnetized model.
Thus, while indications of an inverse cascade are present, they may be
principally due to compressibility.

Note that Figure 3 further indicates that a steady-state saturation has been reached in the spectra for values of $k>k_{peak}$ by time $t=0.05t_{s}$. Similarly in Paper II, Figure 1a shows that total energy saturation of forced models was also reached by time $t=0.05t_{s}$. The characteristic timescale for saturation of hypersonic (${\cal M}_{s}>1$) turbulence is that which characterizes the bulk flow velocity at the driving scale $\lambda_{peak}$, namely the flow crossing time $t_{f}(\lambda_{peak})=t_{f}(L)/8=0.025t_{s}(L)$ for ${\cal M}_{s}=5$.  This is because the relevant shock formation time scale is $t_{f}(\lambda _{peak})$, rather than the sound crossing time $t_{s}$ that would be needed for shock formation in a nonlinear wave steepening process. 

\begin{figure}
  \center \epsscale{1.}
\plotone{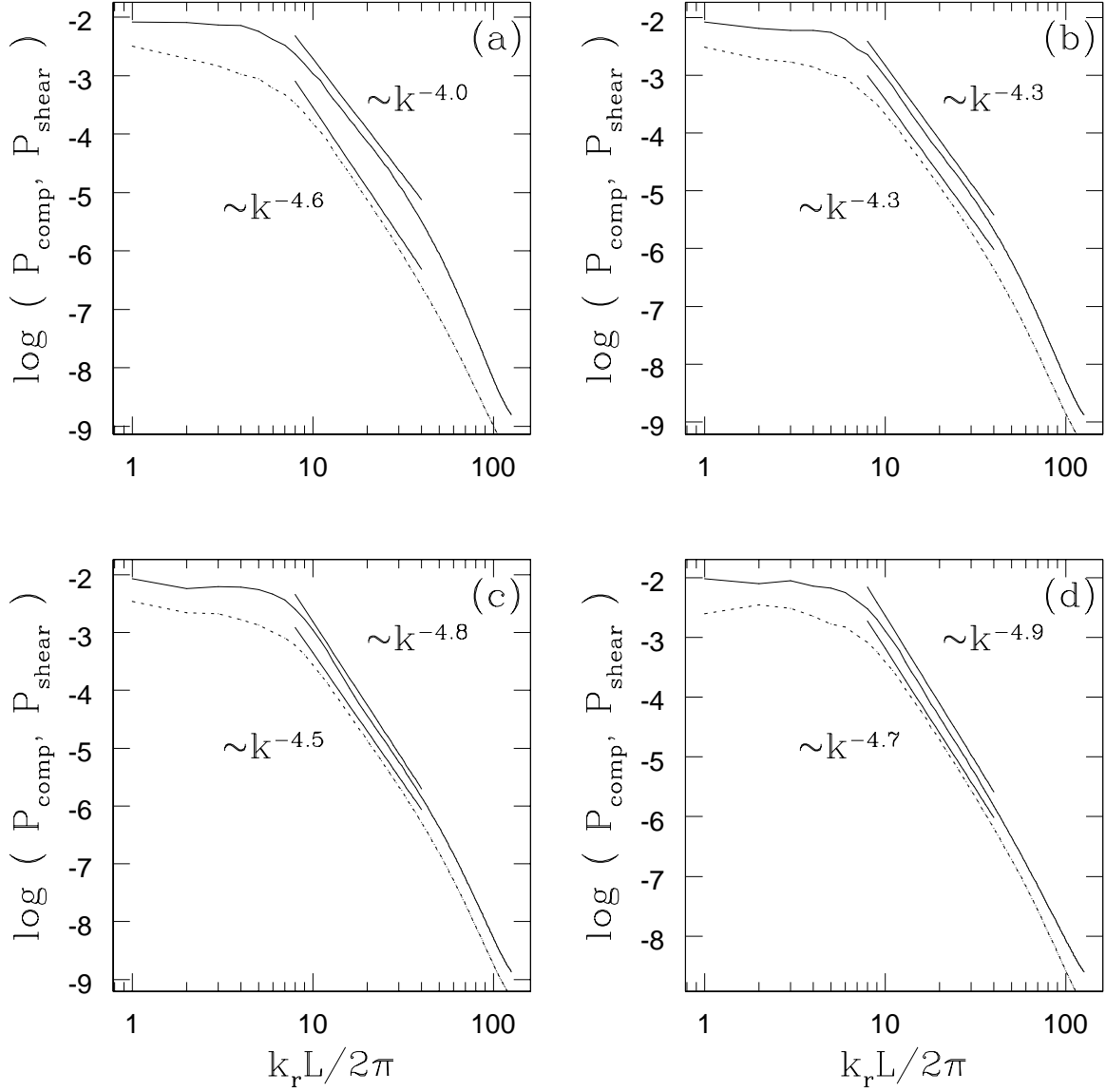}
\caption{Compressive (dashed curves) and shear (solid curves) 
  velocity power spectra for models (a) $\beta=0.01$, (b) $\beta=0.1$,
  (c) $\beta=1.0$, (d) $\beta=\infty$.  Fits are made to the inertial
  range (solid line), as labeled.  The corresponding total percentages of
  compressive energy ($E_{comp}/E_{\rm K}$) are 12\%, 19\%, 21\%, and 24\% for 
  (a)-(d).
\label{fig:plot3}}
\end{figure}

How important is the compressive component of the velocity field in
the dynamics of turbulence? At moderate Mach numbers, it depends on
the ambient magnetic field strength. In Figure \ref{fig:plot3}, we
show an overlay of $P_{shear}(k)$ and $P_{comp}(k)$ for several models of
differing field strength, illustrating the difference in magnitude and
slope of the shear versus compressible components of the turbulent
velocity spectra. Those shown in Figure \ref{fig:plot3} are the lower resolution $256^3$ models, averaged over times $t=0.25t_s$ and $0.30t_s$.

Table 2 provides for several comparisons between models. 
For the strong field case (models A2, A), the compressive energy is an
order of magnitude smaller than the shear energy, and has a steeper slope
(-3.7 vs. -4.3 for the $512^{3}$ models, or -4.0 vs. -4.6 for the
$256^{3}$ models). As the magnetic field strength decreases, the
difference in relative magnitudes and slopes of the shear and
compressive power decreases until, for the pure hydrodynamic model,
the difference in magnitude is about a factor of 3 and the slopes,
-4.7 and -4.9, are almost indistinguishable, given the errors in
average fitting of $P_{comp}(k)$ of model D. The magnitude difference in part 
simply reflects the fact that the forcing is incompressive. 
For weak field cases, the spectral
slopes of the shear and compressive components are also very close (-4.4
and -4.2 for the $512^{3}$ model or -4.7 and -4.5 for the $256^{3}$
model), given the afore mentioned details of fitting averages.

The values we find for the total shear and compressive 
energies for each model, as listed in Table 1, indicate 
that $E_{comp}$ accounts for 24\% of the total kinetic energy 
for the unmagnetized case (model D), 21\% for model C, 19\% for model B, 
and finally only 12\% for the strong field case (model A). 
Comparison of models A, A2; C, C2 shows no dependence on numerical 
resolution of these ratios.

\subsection{Directional dependence of the power spectra}

\begin{figure}
\center
\epsscale{1.}
\plotone{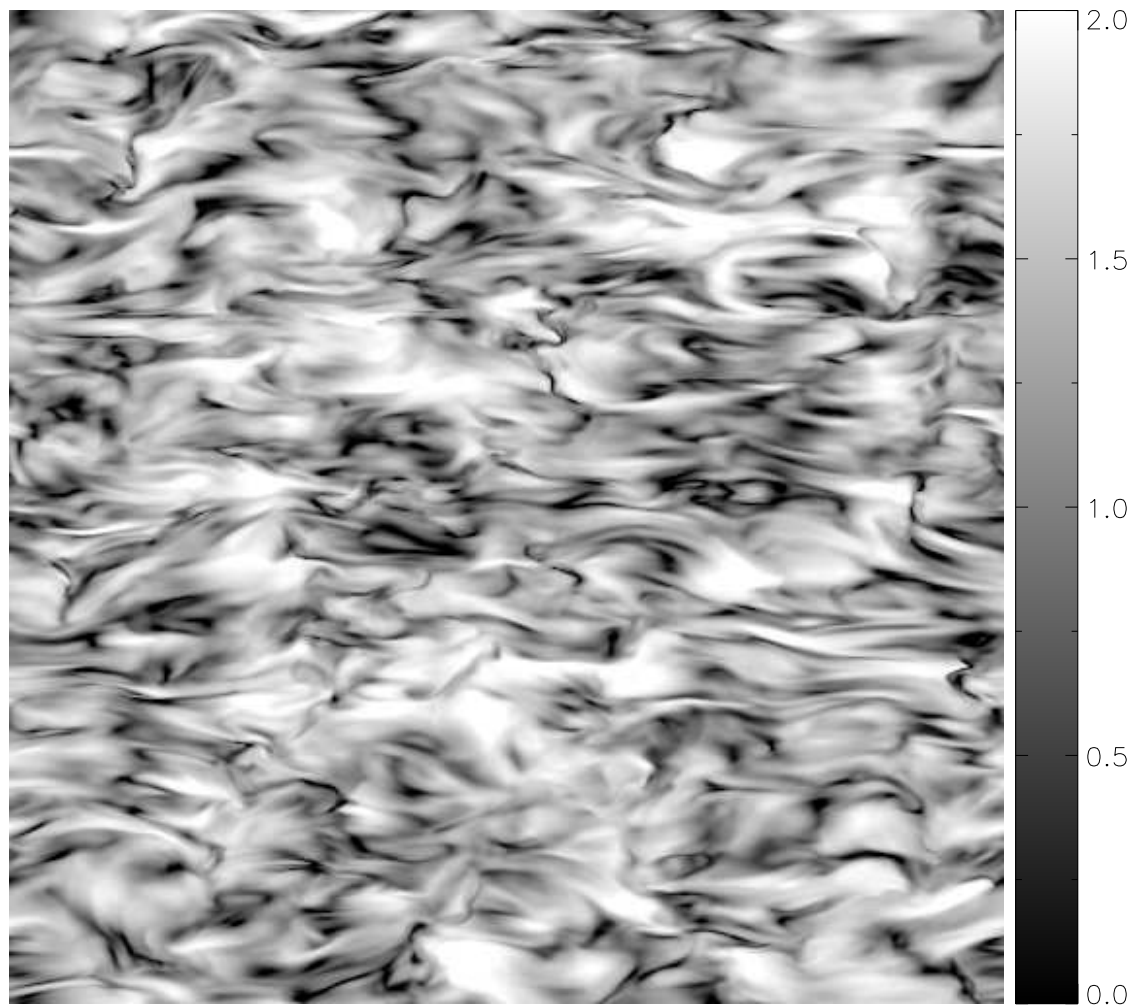}
\caption{Greyscale representation of $(v/c_s)^2$ in a 
slice (z=0) through the computational volume. Anisotropy is evident in the 
elongation of structures in the direction ${\hat x}$ of the mean magnetic 
field (left to right). To emphasize morphological features, $log_{10}$(value) is shown from 0.0 (black) to 2.0 (white).
\label{fig:plot9a}}
\end{figure}

\begin{figure}
\center
\epsscale{1.}
\plotone{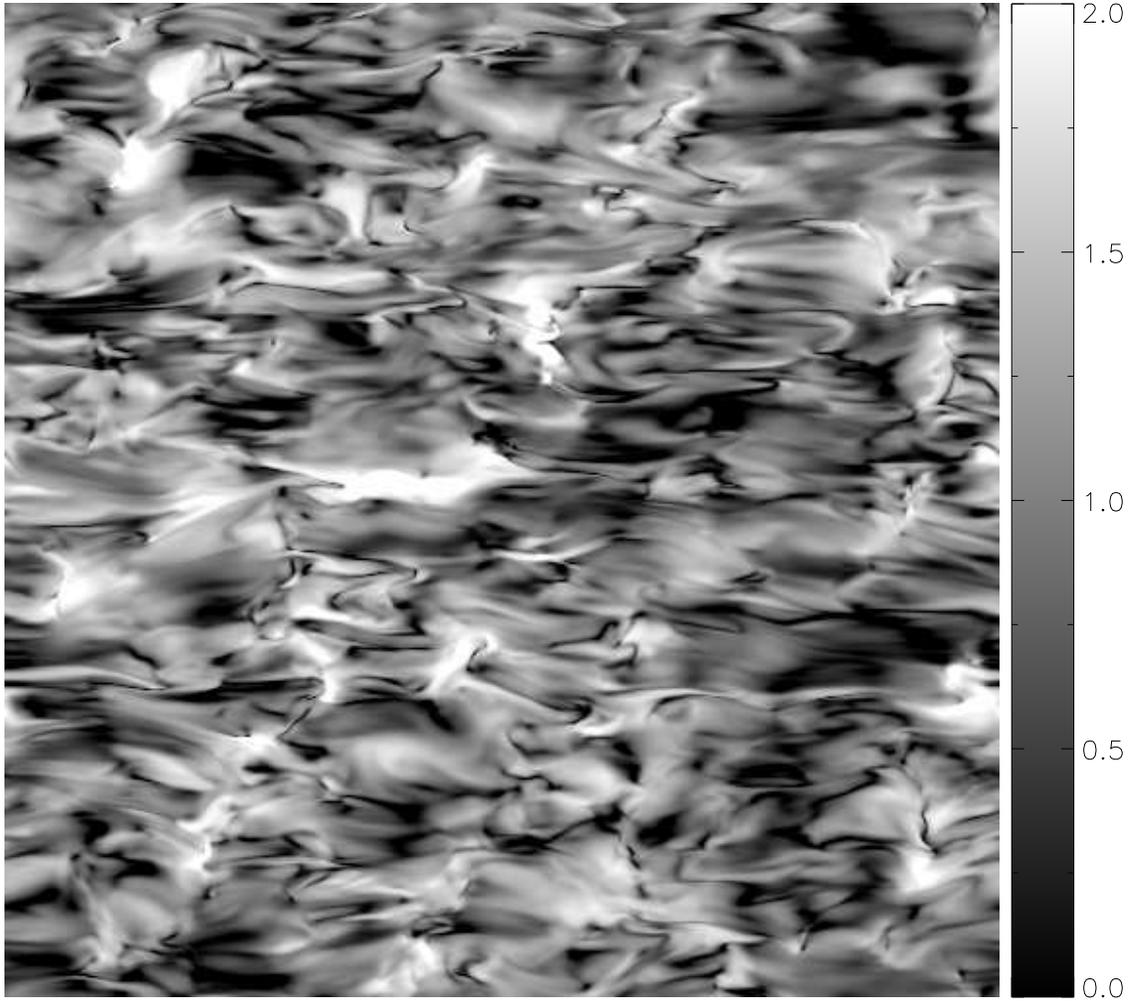}
\caption{Same as in Fig. 5, for $(\delta B)^2/(4\pi\bar\rho c_s^2)$.
\label{fig:plot9b}}
\end{figure}

We begin with a qualitative look at the turbulent anisotropy as seen in
Figures \ref{fig:plot9a}, \ref{fig:plot9b}. These represent gray scale plots of
$(v/c_s)^2$ and $(\delta B)^2/(4\pi \bar\rho c_s^2)$ from 
a 2D slice taken through the
computational box (z=0) for the strong field case ($512^3$ model A2) at time
$t=0.10t_{s}$. Notice that the structures are elongated along the
direction of the mean magnetic field (horizontal axis). There are clearly stronger small scale variations in directions perpendicular to the mean field, suggesting more power in larger ${\mathbf k_{\perp}}$ and in smaller ${\mathbf k_{\parallel}}$.

To study more quantitatively the global spectral anisotropy induced by
the presence of the magnetic field, we have binned the PSD in annular
shells, concentric about the axis of the ambient magnetic field
direction $\hat{x}$, so as to compare the PSD in the plane
perpendicular to the field with that in the direction parallel to the
field. The two-dimensional angle-averaged power spectra thus generated
are functions of $k_{\perp} = (k_{y}^{2} + k_{z}^{2})^{1/2}$ and
$k_{\parallel} = k_{x}$. The 2D contour maps of
PSD($k_{\perp}$,$k_{\parallel}$) provide a visual means by which to
qualitatively characterize the effect of the magnetic field.

We have computed two-dimensional power spectra for all our models, and
find that they become more and more anisotropically distributed with
increasing mean magnetic field strength, with more of the power
concentrated along the directions perpendicular to that of the ambient
magnetic field.

\begin{figure}
  \center \epsscale{1.}
\plotone{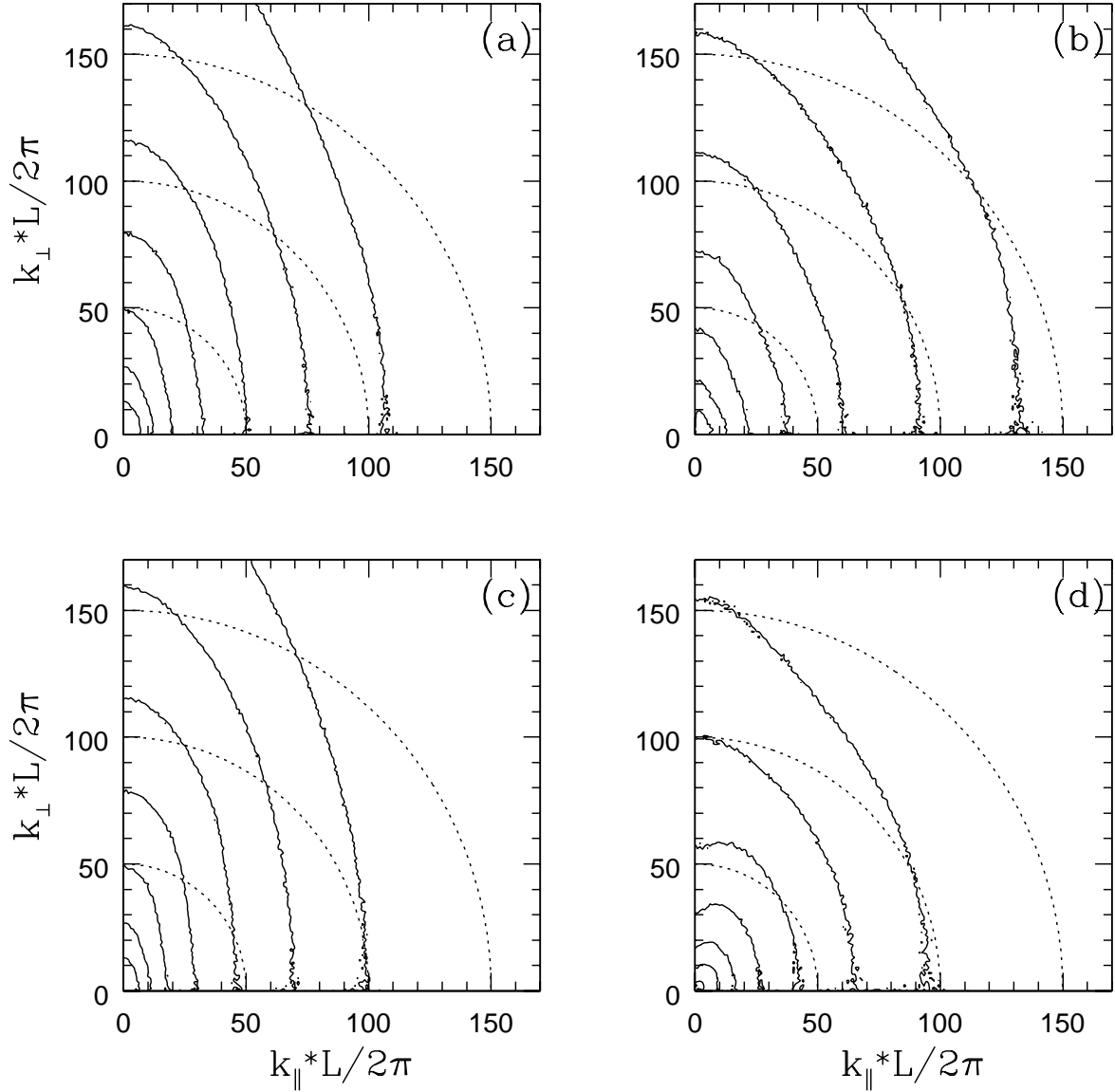}
\caption{Strong field case (model A2, $\beta=0.01$). 
  Contour plots of $P(k_{\perp},k_{\parallel})$ for (a) $P_{\rm K}$,
  (b) $P_{B}$, (c) $P_{shear}$, and (d) $P_{comp}$. Solid contours are
  log($P(k_{\perp},k_{\parallel})$)=-3 to -9, with the largest power being at the smallest
  $k$ (bottom left of each pane). Dotted isotropic curves are included
  for comparison at k=50, 100, and 150.
\label{fig:plot5}}
\end{figure}

Figure \ref{fig:plot5} shows contour plots of power spectra for the
strong field case ($\beta =0.01$, model A2, $t=0.10t_s$). 
In addition to spectra of $P_{\rm K}(k_{\perp},k_{\parallel})$ and $P_{B}(k_{\perp},k_{\parallel})$, we also show the shear $P_{shear}(k_{\perp},k_{\parallel})$ and compressive $P_{comp}(k_{\perp},k_{\parallel})$ velocity power spectra separately. The solid contours represent the levels of constant power, from log(P($k_{\perp}$,$k_{\parallel}$)) = -3 (smallest $k$) to -9 (largest $k$) in steps of 1. Three circular dashed curves at k=50, 100, and 150 are overlayed to illustrate a perfectly isotropic relation between $k_{\perp}$ and $k_{\parallel}$
for comparison.

From Fig. \ref{fig:plot5}, notice that the power remains concentrated
at larger length scales along the magnetic field (small
$k_{\parallel}$) direction, while extending to smaller length scales
in the two perpendicular directions (large $k_{\perp}$). There is also
a clear indication that the anisotropy is dependent on scale size in
the $P_{comp}$ spectra (Fig. \ref{fig:plot5}d).

\begin{figure}
\center
\epsscale{1.}
\plotone{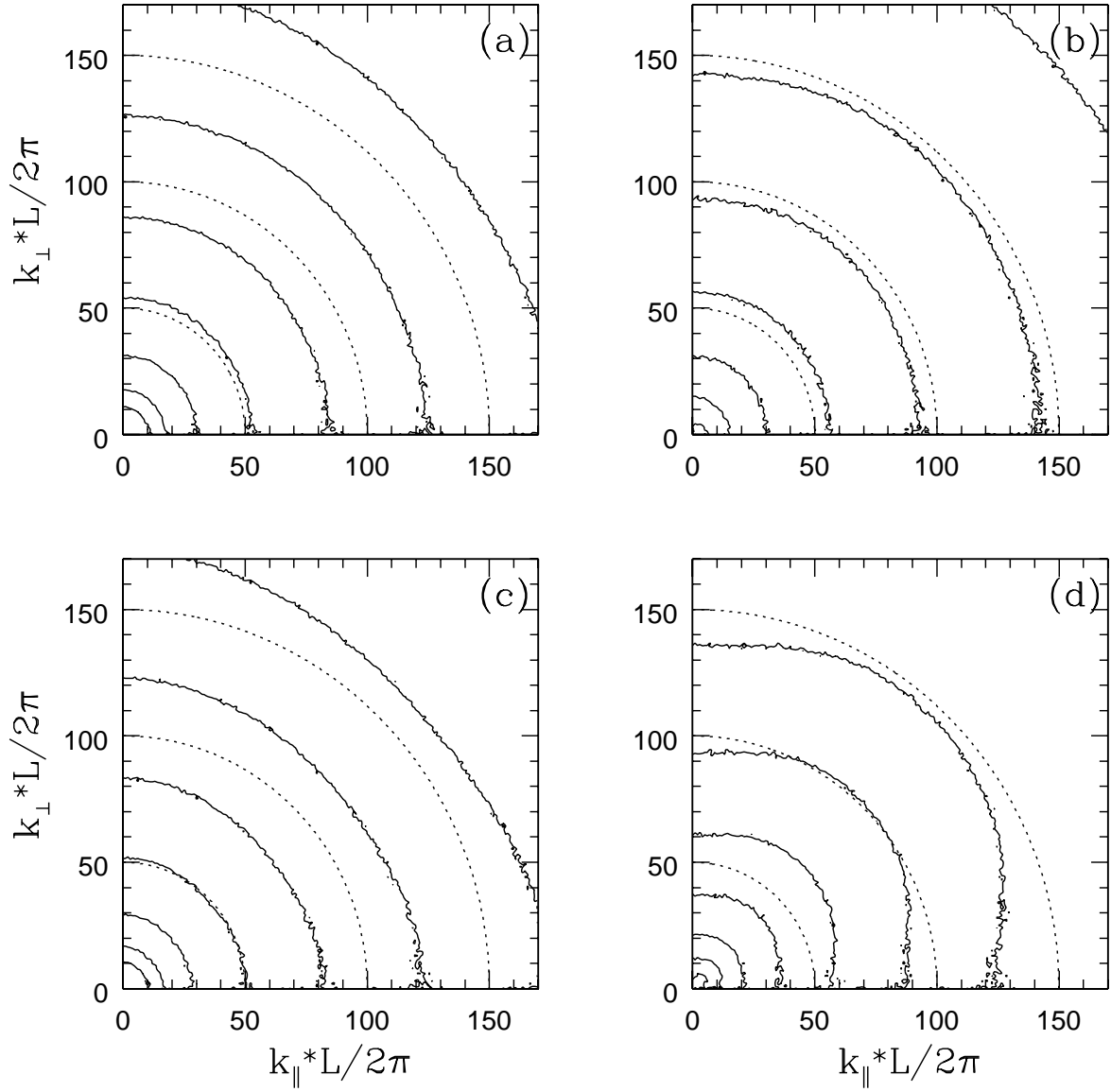}
\caption{Same quantities and levels as in Fig. \ref{fig:plot5}, for 
weak-field case (model C2, $\beta=1.0$).
\label{fig:plot6}}
\end{figure}

Compared to our results for the $\beta =0.01$ model, we find that the
power distributions are dramatically more isotropic with a decrease in
field strength. Figure \ref{fig:plot6} shows contour plots for the
power in the weak field case ($\beta =1.0$, model C2, $t=0.15t_s$). While the
magnetic field and total velocity field are clearly isotropic, there
is a small but real anisotropy along a 45 degree angle between
$k_{\perp}$ and $k_{\parallel}$ for both the shear and compressive
components of velocity. Note how the compressive field (Fig.
\ref{fig:plot6}d) has deficiency in power (extended contour) along the
45 degree line, while the shear field (Fig. \ref{fig:plot6}c) has a
surplus (contracted contour) along the same line; the total velocity
obtained as the sum of these is, however, isotropic (Fig.
\ref{fig:plot6}a).

To further investigate direction-dependent behavior of the power
spectra, we have taken slices through our 2D contour plots along
$k_{\perp}=1$ and $k_{\parallel}=1$ to generate 1D plots of
$P(k_{\parallel})$ and $P(k_{\perp})$, respectively. For all our
models, the qualitative morphological features of the directional
$P(k_{\perp})$ and $P(k_{\parallel})$ spectra include driving,
inertial, and dissipative ranges similar to the angle-averaged PSD
curves already shown in Figures \ref{fig:plot1} and \ref{fig:plot2}.

\begin{figure}
\center
\epsscale{1.}
\plotone{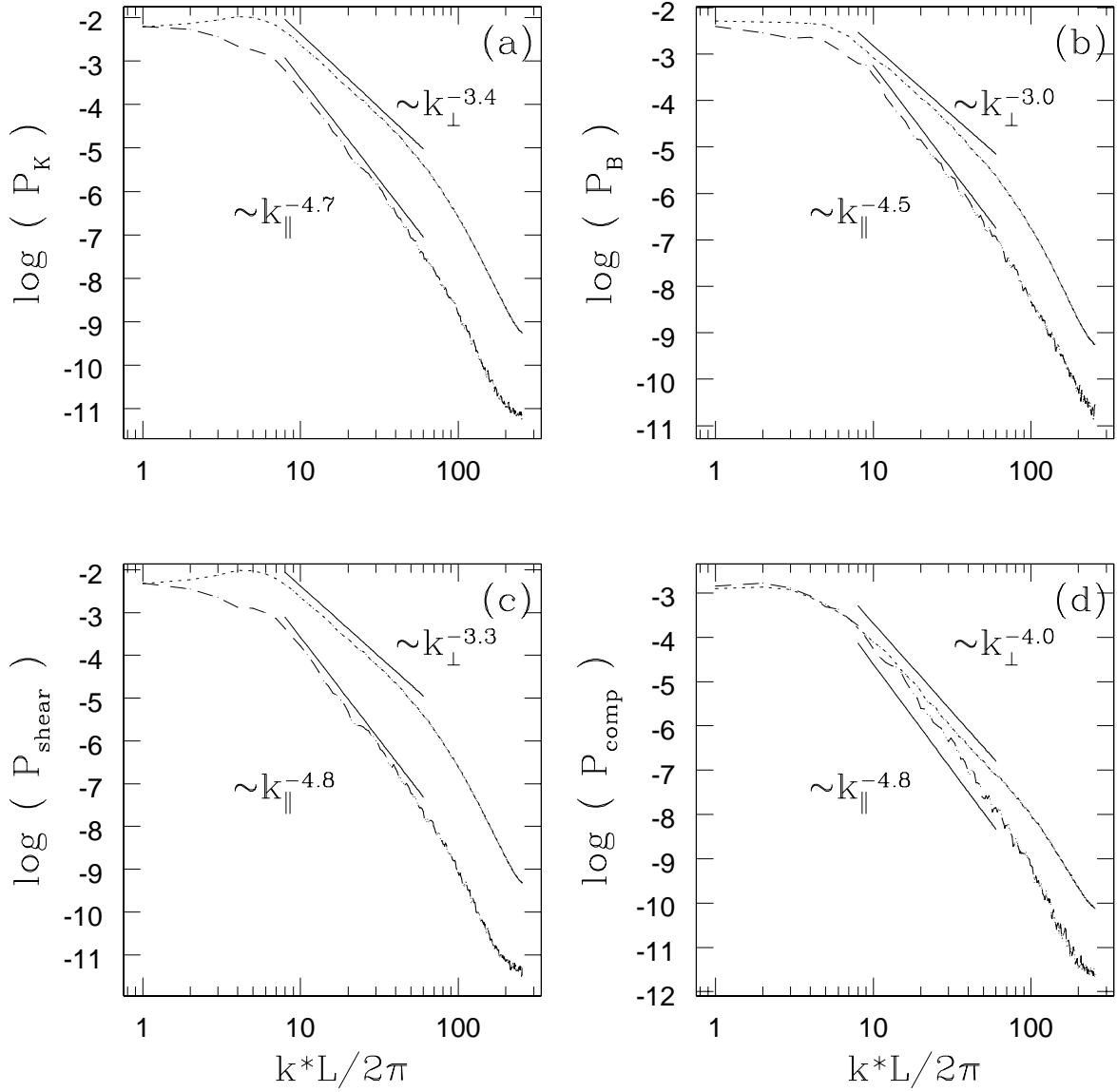}
\caption{Strong field case ($\beta=0.01$, model A2). 
  Overlays of P($k_{\parallel}$) (dashed line) and P($k_{\perp}$)
  (dotted line) for (a) $P_{\rm K}$, (b) $P_{B}$, (c) $P_{shear}$, and (d)
  $P_{comp}$. Fits are made to the inertial range (solid lines).
\label{fig:plot7}}
\end{figure}

For the strong-field model, there are clear differences in the
directional power that reflect the spectral anisotropy evident in
Figure \ref{fig:plot5}; we illustrate the results in Figure \ref{fig:plot7}.
From Figure \ref{fig:plot7}, spectral anisotropy, as represented in
the difference in magnitude of the overlayed $P(k_{\perp})$ and
$P(k_{\parallel})$ curves, appears strongest in the cases of $P_{\rm K}$ 
and $P_{shear}$, with the scale dependence of this anisotropy clearly
evident in the differing slopes characterizing the inertial ranges.
The magnitude of the power with variations 
in the direction parallel to $\bf{B}$ is
comparable to that of the power with variations 
perpendicular to the field only for
the largest length scales. As $k$ increases, $P(k_{\parallel})$
falls off much faster than $P(k_{\perp})$, such that the anisotropy
increases with decreasing length scale.

Within the driving range, the incompressible nature of the driving is
clearly evident from the directional spectra in Figure
\ref{fig:plot7}, as seen in by the hump near $k_{peak}$ which appears
strongly in the $P(k_{\perp})$ curve for $v_{shear}$, but not at all
in the curves for $v_{comp}$.  Within the inertial range, both curves
roughly follow a power law. It is difficult to identify a clear
separation between the inertial and dissipative ranges for the
$P(k_{\parallel})$ curves. Note that from Figure \ref{fig:plot7}d, for
the largest scales there is isotropy in the distribution of compressive
velocities, but as $k$ increases, $P_{comp}(k_{\perp})$ becomes larger than
$P_{comp}(k_{\parallel})$.

To construct quantitative measures of the anisotropy and to see how
these vary with changing magnetic field strength, we can compare
indices of power-law fits through the inertial range of each of these
$P(k_{\perp})$ and $P(k_{\parallel})$ curves, as done before with the
angle-averaged power spectra. Table 3 shows the index $n$ of the
$k^{-n}$ power-law fits for $P_{B}$, $P_{\rm K}$, $P_{shear}$, and
$P_{comp}$ for all models; in Figure \ref{fig:plot7}, these fits are
compared to the directional spectra for model A2. Note that such inertial range
fits are only able to show an anisotropy in the strong field models.
The previously mentioned subtle variations in the weak field models
$P_{comp}$ and $P_{shear}$ (Fig. \ref{fig:plot6}) are not seen here.

There is a general trend that the $P(k_{\perp})$ curves for the total
and component velocity power spectra exhibit a steeper slope for the
weak field than the moderate field case, and more so for the
hydrodynamic case. This is especially seen for $P_{K}(k_{\perp})$, 
where the indices of the power-law fits are 3.5, 4.3,
4.7, and 4.8 for the $\beta =$ 0.01, 0.1, 1.0, and $\infty$ models
respectively. Note however that the reverse trend is evident for
$P_{B}(k_{\parallel})$, with spectral indices of 4.5, 4.0, 3.6 for
the $\beta =$ 0.01, 0.1, and 1.0 models respectively.

The strong field in models A and A2 is responsible for the
anisotropy in the power spectra as reflected in the relative slopes along
$k_\parallel$ and $k_\perp$ directions.  The steepest profiles for any 
component of our models are for the
strong field model along the direction parallel to the field, while
the slope in the perpendicular direction for the same model is shallowest. 
For $\beta = 0.01$ (model A2), $P_{B}(k_{\parallel}) \propto
k_{\parallel}^{-4.5}$ while $P_{B}(k_{\perp}) \propto
k_{\perp}^{-3.0}$. For the same model, $P_{K}(k_{\parallel}) \propto
k_{\parallel}^{-4.7}$ and $P_{K}(k_{\perp}) \propto k_{\perp}^{-3.4}$.
This is indicative of the fact that increasingly small amounts of
power are transfered into the $k_{\parallel}$ direction at large $k$.

\begin{figure}
  \center \epsscale{1.}
\plotone{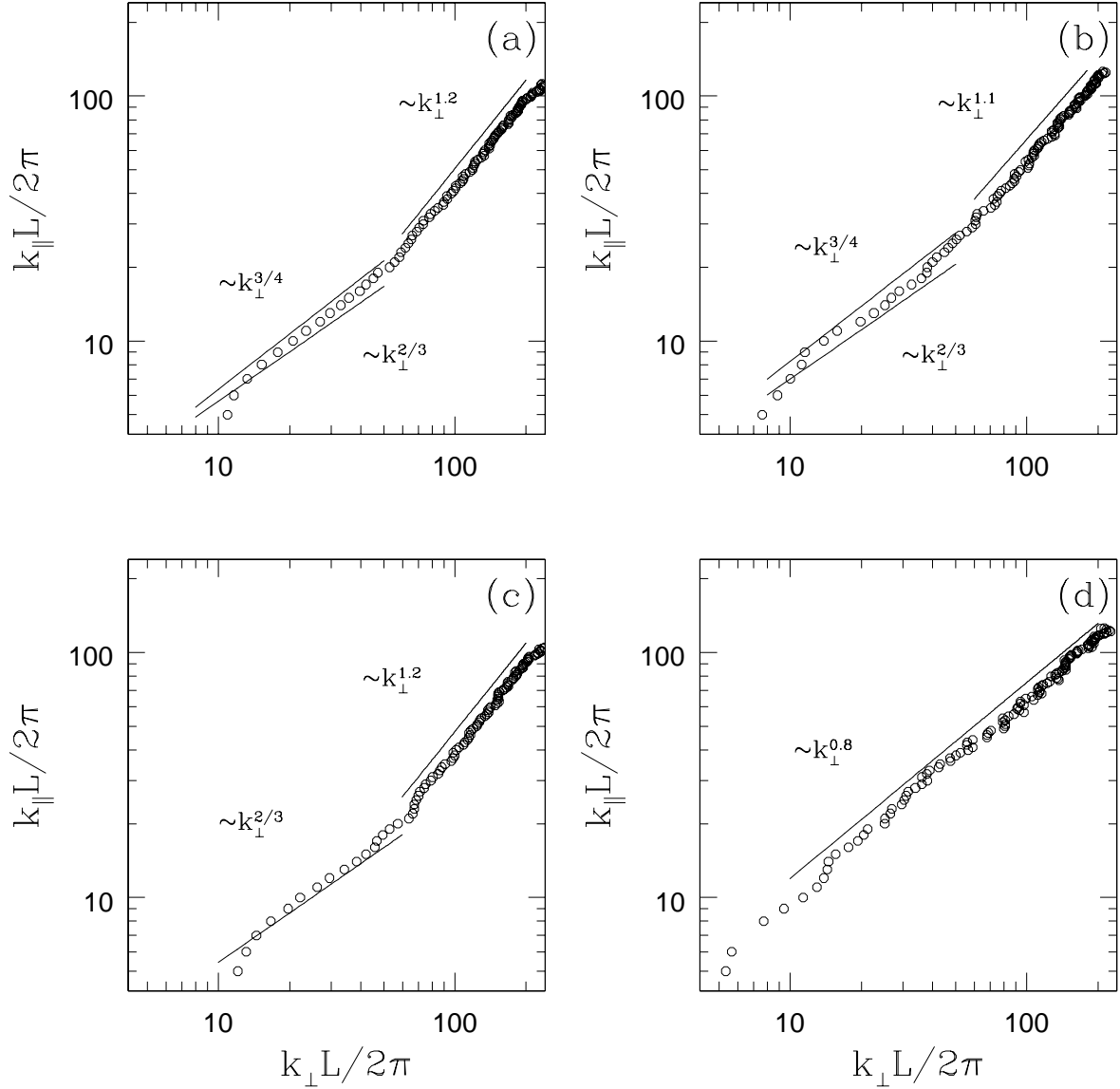}
\caption{Measure of global spectral anisotropy for the strong field case 
  ($\beta=0.01$, model A2).  Plots of $k_{\parallel}$ vs. $k_{\perp}$
  for (a) $P_{\rm K}$ , (b) $P_{B}$, (c) $P_{shear}$, and (d) $P_{comp}$.
  Points of $k_{\parallel}$-intercept versus the $k_{\perp}$-intercept
  for a given power contour, as interpolated from the P($k_{\perp}$)
  and P($k_{\parallel}$) curves. 
\label{fig:plot8}}
\end{figure}

As we shall discuss further in \S 5, the exponents of the power-law
fits to $P(k_{\perp})$ and $P(k_{\parallel})$ (for both $P_{\rm K}$
and $P_{B}$) can be related to a power-law scaling behavior between
$k_{\parallel}$ and $k_{\perp}$ that describes the spectral anisotropy
(cf. GS).  This can be quantified (see \cite{Cho2000}) by asking, 
for a given power contour, what are the $k_{\perp}$ and 
$k_{\parallel}$ intercepts?
Starting with our 2D spectra, we compute a $k_{\parallel}$-$k_{\perp}$
relation by performing an interpolation of the P($k_{\parallel}$)
curve into the P($k_{\perp}$) curve. The resulting function
$k_{\parallel}$($k_{\perp}$) provides information on how $k_{\perp}$
and $k_{\parallel}$ are related across a given range of power
contours.  This scaling is not constant across all magnitudes of
power, but is relatively constant within the inertial range. Figure
\ref{fig:plot8} shows plots of $k_{\parallel}$($k_{\perp}$) from the
strong field model for $P_{B}$, $P_{\rm K}$, $P_{shear}$, and $P_{comp}$.
In the inertial range, the relation for $P_{B}$, $P_{\rm K}$, and
$P_{shear}$ is consistent with $k_{\parallel} \propto k_{\perp}^{2/3}$
or $k_{\perp}^{3/4}$; the compressive velocity has quite different
behavior, roughly $ k_{\parallel} \propto k_{\perp}^{0.8}$. The
scaling behavior for $P_{comp}$ holds over an extensive range, both at
small $k$ and for $k$ within the expected dissipative range.

We conclude this section with a discussion of one possible application of this work to observations. Namely, we propose a method for constraining the value of the mean magnetic field strength within observed cloud cores by analyzing the degree of anisotropy in centroid maps in molecular tracers. Consider a cloud core which has a nonzero component of the mean magnetic field lying in the plane of the sky ($xy$-plane) and a relatively uniform density $\rho$ (for the later to be true, the velocities must be sonic or below). Given spatially uniform molecular excitation, the centroid of an emission line at a position $(x,y)$ in a projected map represents the mean value of the line-of-sight turbulent velocity,  $\langle v_{z}\rangle (x,y)$=$\int v_{z}(x,y,z) dz/\int dz$. If one takes the Fourier transform of the 2D velocity 
centroid map, the result $\langle v_{z}\rangle (k_{x},k_{y})$ is identical to taking the $k_{z}=0$ plane from the Fourier transform $v_{z}(k_{x},k_{y},k_{z})$ of the full 3D data cube. In Figure 11a and 11b, we compare the results of these two equivalent calculations, i.e. the FFT of the ``centroid map'' $\langle v_{z}\rangle (x,y)$ and the $k_z=0$ slice taken from the FFT of the full 3D data cube of $v_{z}$ for the strong-field model A. The same anisotropy previously discussed, albeit with greater noise, is evident in projected maps. 
\footnote{We note, however, that anisotropy in velocity centroid maps can be washed out if the emissivity is highly nonuniform, as expected for any region having supersonic flow. Direct tests on our ${\cal M}=5$ models show, for example, that {\it density-weighted} velocity centroid maps have isotropic power spectra.}

In Figure 11c and 11d, we show that $\langle v_{z}\rangle (k_{x},k_{y})$ from the weaker-field models (B,C) has significantly lower anisotropy than does model A. Taken together, these results suggest that anisotropy in the Fourier transform of velocity centroid maps can provide a lower bound on the mean magnetic field strength in an observed cloud. From Table 1, ${\cal M}_{A,0}=\langle v^{2}\rangle^{1/2}/v_{A,0}$ in model A is 0.5, while ${\cal M}_{A,0}$ in model B is 1.5. Here $v_{A,0}$ is the Alfv\'{e}n speed in the mean field, which for projections along ${\hat z}$ is also the mean plane-of-the-sky field, $v_{A,p}$. Allowing for a factor of $1/{\sqrt 3}$ reduction for $\langle v_{z}^{2}\rangle^{1/2}$ compared to $\langle v^{2}\rangle^{1/2}$, this implies that when $\langle v_{z}^{2}\rangle^{1/2}/v_{A,p}$ is $\sim 0.3$, anisotropy would be clearly evident (unless washed out by other effects such as strong but isotropic density variations), whereas when $\langle v_{z}^{2}\rangle^{1/2}/v_{A,p}$ is $\sim 0.9$ or larger, negligible anisotropy is expected.

\section{Summary and discussion}

In this paper we present spectral analysis of driven, highly
compressive (${\cal M}_{s}\sim 5$) MHD turbulence from a collection of 
3D numerical simulations whose parameters span a range believed to
apply within GMCs.  We investigate models that were the subject of
previous work on dissipation of turbulence (Paper II), as well as new
high resolution counterparts. Each of our models evolves an isothermal
plasma in time subject to an applied incompressive turbulent driving
of the velocity field. We parameterize the strength of the mean magnetic
field $B_{0}$ by $\beta=c_{s}^{2}/v_{A,0}^{2}$ and compare models of varying
$\beta$. Our main findings, and their relation to other work, are outlined
as follows:

1. The angle-averaged spectra for the total velocity and its shear and
compressive components ($P_{\rm K}$, $P_{shear}$, $P_{comp}$) all
exhibit distinctive environmental, driving, inertial, and dissipative
ranges.  In some of the spectra, there is evidence within the inertial
range (i.e. for wavenumbers not directly populated by forcing) of a
slight break. In these cases, the portion to the left of the break may
represent a transition region, and the portion to the right an
asymptotic regime. For strong-field models, the spectrum of $P_B$ also
shows these distinct ranges. The spectra for $P_B$ shows no inertial
range for the weaker field models, while spectra of $\rho$ show none
for any model.

2. Signs of a nonzero but weak inverse cascade toward modal energy
equipartition (i.e. flat $P(k)$) is evident for both magnetized and
unmagnetized models at $k<k_{peak}$. This develops within half a flow crossing time over the scale of the box ($0.1t_{s}$, see Fig. 3), corresponding to four crossing times at the driving scale $\lambda_{peak}$. 
Observations of the turbulent spectra in clouds do not show such 
flattening towards large scales.  
Our work thus places serious constraints on the importance of 
turbulent driving at length scales ${\ell }\ll L$ within GMCs,
in comparison to inheritance of turbulence with an extended power spectrum
from the process of cloud formation.

3. For the strong field case (model A2), the values of the index $n$
for the spectra of $P_{turb}=(P_{\rm K}+P_{B})$, $P_{\rm K}$, and $P_{shear}$
are in the range $3.7-3.8$, while the value for $P_{B}$ is slightly
less steep, $\approx 3.5$.  The velocity spectra averaged over the inertial range are thus consistent
with the strong MHD turbulence model of GS, which has $n=11/3$.  In recent
numerical work for incompressible MHD, \cite{Cho2000} reported a 1D
spectrum consistent with $n=11/3$, as did \cite{Muller2000} and
\cite{Biskamp2000}. \footnote{Both groups note that numerical schemes
  employing hyperviscosity show spectral flattening at large $k$ from
  the so-called bottleneck effect; because ZEUS uses
  finite-differencing rather than a spectral method, our results do
  not show this bottleneck.}  \cite{MG2001} report a shallower law
consistent with $n=7/2$ for their angle-averaged energy spectra; although
they regard this as anomalous due to its inconsistency with GS's
theory, we find it intriguing that our magnetic power spectra are
similar.  Interestingly, when we fit just the highest-$k$ portion of the inertial range, we also find $n_{r}=3.5$ for $P_{K}$ and $P_{shear}$ for model A2. \cite{Cho2002} also report a slope similar to $11/3$ for
Alfv\'en modes in compressible MHD simulations, although their spectra
are somewhat noisy because the inertial range falls at relatively low
$k$ in their models.

4. For the weak field case (model C2), values $n$ of the spectral
index for $P_{turb}$, $P_{\rm K}$, $P_{shear}$, and $P_{comp}$ are in
the range $4.0-4.4$, steeper than $11/3$ and even exceeding the index
$4$ associated with Burgers's spectrum. Although these measurements
may be affected in part by spectral steepening toward the driving
range, the slopes to the right of the ``break'' in the inertial range
(e.g. 3.9 for $P_{K}$) remain steeper than those in model A2 (e.g. 3.5
for $P_{K}$).  Thus, a cloud permeated by a weak {\it mean} magnetic
field -- even if the {\it total} magnetic field strength is
significant (cf. the total magnetic energy in Table 1) -- would be
expected to yield a distinctly steeper velocity spectrum than a cloud
with a stronger mean field.  $P_B$ for the weak-field model has a much
shallower nominal spectral slope, $3.3$, but with large uncertainties
due to fitting a limited inertial range.

5. As the mean magnetic field strength decreases, the slopes of the
spectra for $P_{turb}$, $P_{\rm K}$, and $P_{shear}$ all become steeper,
implying that a stronger magnetic field inhibits dissipation somewhat
at intermediate length scales.  For a weak-field system, velocities
are super-Alfv\'enic (leading to shock dissipation) over a larger
range of scales than in a stronger-field system, so that the
near-incompressive limit for which the GS model applies would be
shifted to larger $k$.  More generally, compressional effects become
progressively stronger for decreasing field strength: the ratio of
compressive energy $E_{comp}/E_{\rm K}$=0.12. 0.19, 0.21, and 0.24 for
models of $\beta=0.01$, 0.1, 1.0, and $\infty$, respectively. 

In part, the lower proportion of compressive energy in the
strong-field models must reflect the tendency for energy injected into
the Alfv\'enic cascade (which is purely shear) to remain there
(cf. GS, \cite{MG2001}, \cite{Cho2002}).  Because losses from compressive 
components are more rapid than from shear components, however, the 
steady-state ratio $E_{comp}/E_{total}$ may differ from the relative losses
through shocks vs. other dissipation (e.g. model A2 loses $\sim 23\%$ of 
the input power in shocks, while $E_{comp}/E_{total}=0.07$)

6. Two-dimensional power spectra $P(k_\parallel,k_\perp)$ in our
strong-field model indicate global anisotropy, with spectral energy
preferentially aligned along $k_{\perp}$  for $P_{B}$, $P_{\rm K}$,
$P_{shear}$, and $P_{comp}$.  Our weak-field models, however, do not show
global anisotropy in their power spectra (but see below).  These findings
are consistent with the results of \cite{MOGH98} and \cite{OMG98}
that global spectral anisotropy (as measured using the mean anisotropy angle
$\theta_\omega$) decreases as the ratio of fluctuating to mean magnetic field
increases.

7. Enhancement of anisotropy at large $k$ is clearly evident in the 
two-dimensional power spectra $P(k_{\perp},k_{\parallel})$, as well as 
in the cuts $P(k_\perp)$ and $P(k_\parallel)$, for our strong-field model.
The global spectral anisotropy of $P_{turb}$, $P_{B}$, and $P_{shear}$ in 
strong-field models is consistent with the relation 
$k_\parallel \propto k_\perp^{2/3}$ proposed by GS, while for 
$P_{comp}$ we find $k_\parallel \propto k_\perp^{4/5}$.

The {\it global} spectral anisotropy scaling we identify in our strong-field 
model has previously been shown to describe {\it local} spectral anisotropy for
simulated incompressible MHD turbulence by \cite{Cho2000} and \cite{MG2001}.
As those works emphasized, the anisotropy predicted by GS is for $\hat k$ 
directions defined with respect to the {\it local} magnetic field.  If the 
large-scale field is strong in the sense that the turbulence is sub-Alfv\'enic
(as for our models A, A2), then the local directions of field lines 
do not depart strongly from $\hat B_0$;  for weak large-scale fields, on the
other hand, local directions of $\hat B$ are uncorrelated with $\hat B_0$
(see Fig. 2 of Paper II).  Thus, our finding of global spectral 
anisotropy only in our strong-field models is consistent with expectations, 
since only for models A, A2 does $\hat B_0$ correlate well with the local 
field.  

Discerning small-scale local anisotropy even when turbulence is
super-Alfv\'enic on large scales is possible using second-order directional
structure functions for coordinates defined with respect to the local magnetic 
field \citep{Cho2000,MG2001}.  Although we have not applied these specialized 
techniques here, our results for weak-field models are also consistent 
with the findings of \cite{Milano2001} that global spectral measures may 
be isotropic even when local anisotropy is present as identified 
by other methods.  

8. The dependence of global velocity spectral anisotropy on the mean
magnetic field strength suggests a potentially important observational
diagnostic. Namely, observational evidence of turbulent anisotropy at
a given spatial scale would imply that the value of the ``mean field''
Alfv\'en Mach number -- i.e. based on the plane-of-the-sky
$|\langle{\bf B}\rangle|$ for that spatial scale -- is less than
unity.  In molecular cloud core regions where the density varies
weakly in space, Fourier transforms of velocity centroid maps
correspond directly to slices through the 3D velocity power spectrum,
so evidence of anisotropy is immediately accessible (see Fig. 11). In
regions where the flow is supersonic, convolution of the anisotropic
velocity power spectrum with the isotropic density power spectrum
tends to wash out anisotropy in velocity centroid maps, but more
complex analysis techniques using the full position-velocity data
cubes may still be able to ascertain the degree of anisotropy in the
velocity power spectrum itself.  Further investigation will show if
such spectral diagnostics offer a practical indirect means of
constraining the magnetic flux through a cloud core, and hence in
determining whether it is magnetically sub- or super-critical with
respect to gravitational collapse.

9. Finally, we briefly comment on the relationship of our results to
proposals presented in \cite{Boldyrev2002a} and \cite{Boldyrev2002b}
 that the spectral slope for
compressible MHD turbulence may be derived from a modified version of
a She-L\'{e}v\^{e}que analysis that provides intermittency corrections to
Kolmogorov scalings (see \cite{SL94}; \cite{Dubrulle94}).  
Those works suggest that for a range of fractal 
dimensions for the dissipative structures,
the index of the power spectrum would lie in the range $3.74-3.89$.
For our fits over the whole inertial range, 
we find indices for $P_{K}$ of 3.8 and 4.3, respectively, for models A2
and C2, and indices 3.7 and 4.4 for just the shear component of
velocity $P_{shear}$. Fitting just the high-$k$ portion of the inertial range for model A2 (C2), we find $n_{r}$ of 3.5 (3.9) for $P_{K}$ and 3.5 (4.0) for $P_{shear}$. Thus, while our results for the strong-field
model A2 are potentially consistent with the Boldyrev et al proposal, the
spectra of our weak-field model C2 are steeper.  
While detailed discussion of the reason for this difference 
is beyond the scope of this paper, we note that the 
basic assumption of a conservative cascade (implicit in the
assumption that the slope of the third-order structure function 
$\zeta(3)=1$) is more applicable in the strong-field model that the weak-field
model. It remains an open challenge to develop an analytic model that
incorporates both direct shock dissipation and multi-scale energy cascades
in a single framework.

\acknowledgements{We are grateful to Charles Gammie for helping to initiate this project and to the referee for a valuable and careful report. This work was supported in part by NASA grants NAG53840 and NAG59167.}

\clearpage

\begin{figure}
  \center \epsscale{1.}
\plotone{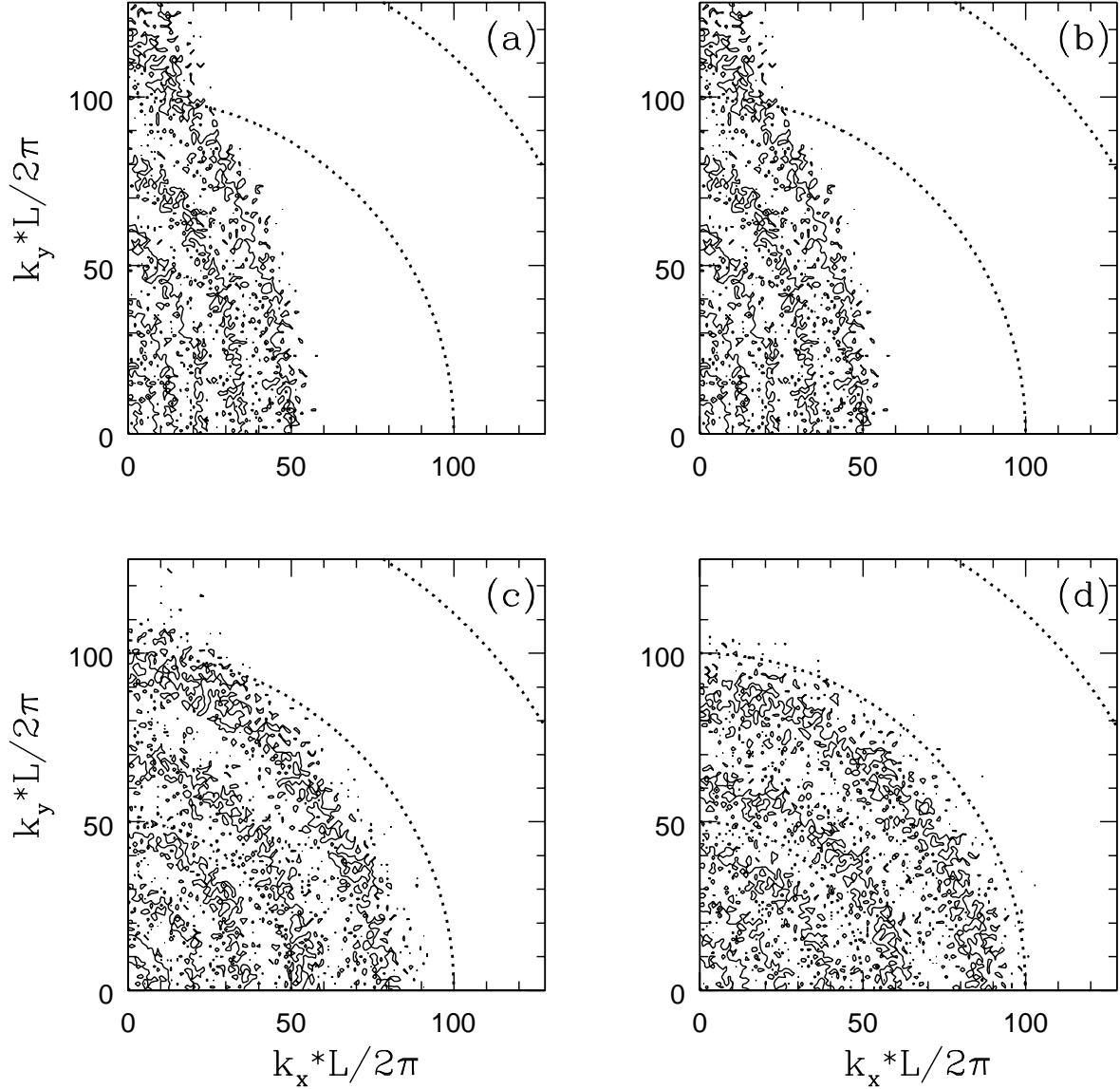}
\caption{Spectral anisotropy in simulated observations as a constraint for magnetic field strength. A comparison of plots for (a) $v_{z}(k_{x},k_{y},k_{z}=0)$ and (b) $\langle v_{z}\rangle (k_{x},k_{y})$ for the strong field model A show that velocity centroid maps with uniform-density conditions evidence the same anisotropy as the underlying power spectrum. Much lower anisotropy in observations of $\langle v_{z}\rangle (k_{x},k_{y})$ are expected for conditions similar to the (c) moderate-field model B, and (d) the weak-field model C.
\label{fig:plot11}}
\end{figure}

\clearpage

\begin{deluxetable}{crrrrrrrrrrr}
\tabletypesize{\scriptsize}
\tablecaption{Model Parameters \label{table1}}
\tablewidth{0pt}
\tablehead{
\colhead{Model} & 
\colhead{$\beta$} & 
\colhead{N} &
\colhead{${\cal M}_{s}$} & 
\colhead{${\cal M}_{A,0}$} & 
\colhead{${\cal M}_{A,rms}$} &
\colhead{$E_{\rm K}$} &
\colhead{$\delta E_{B}$} &
\colhead{$E_{shear}$} &
\colhead{$E_{comp}$} &

}
\startdata
A  &0.01 &256 &5.11 &0.511 &0.477 &13.1 &7.35 &11.5  &1.55 \\
B  &0.1  &256 &4.86 &1.54 &0.986 &11.8 &7.14 &9.60  &2.20 \\
C  &1.0  &256 &5.08 &5.08 &1.67 &12.9 &4.13 &10.1  &2.77 \\
D  &$\infty$ &256 &5.56 &N/A &N/A &15.5 &N/A  &11.8  &3.71 \\
$C_{k4}$ &1.0 &256 &6.61 &6.61 &1.75 &21.9 &6.42  &17.4  &4.76 \\
A2 &0.01 &512 &5.26 &0.526 &0.490 &13.8 &7.65  &12.4  &1.44 \\
C2 &1.0  &512 &5.16 &5.16 &1.55 &13.3 &5.05  &10.5  &2.81 \\
\enddata

\tablecomments{Energies in units $\rho L^{3} c_{s}^2$.}

\end{deluxetable}

\clearpage

\begin{deluxetable}{crrrrrrrrrrr}
\tabletypesize{\scriptsize}
\tablecaption{Comparative Spectral Slopes  \label{table2}}
\tablewidth{0pt}
\tablehead{
\colhead{Model} & 
\colhead{$\beta$} & 
\colhead{$P_{turb}(k)$} &
\colhead{$P_{B}(k)$} & 
\colhead{$P_{\rm K}(k)$} & 
\colhead{$P_{shear}(k)$} & 
\colhead{$P_{comp}(k)$} &
}
\startdata
A  &0.01     &3.9  &3.7  &4.0  &4.0  &4.6  \\
B  &0.10     &4.0  &3.6  &4.3  &4.3  &4.5 \\
C  &1.00     &4.3  &3.5  &4.7  &4.7  &4.5 \\
D  &$\infty$ &4.8  &N/A  &4.8  &4.9  &4.7 \\
A2 &0.01     &3.7  &3.6  &3.8  &3.7  &4.3  \\
C2 &1.00     &4.0  &3.3  &4.3  &4.4  &4.2  \\
\enddata 
\tablecomments{Exponents $n$ for the power-law fits $P(k)\propto
  k^{-n}$ of energy component power spectra (PSD). Error estimates for the fits are discussed in the main text. Note that the values of $n$ 
corresponding to a Kolmogorov and Burgers spectra are 11/3 and 4 respectively.
}
\end{deluxetable}

\clearpage

\begin{deluxetable}{crrrrrrrrrrr}
\tabletypesize{\scriptsize}
\tablecaption{Comparative Spectral Slopes of Directional Spectra \label{table3}}
\tablewidth{0pt}
\tablehead{
\colhead{Model} & 
\colhead{$\beta$} & 
\colhead{$P_{B}(k_{\parallel})$} &
\colhead{$P_{B}({k_{\perp}})$} & 
\colhead{$P_{\rm K}(k_{\parallel})$} &
\colhead{$P_{\rm K}({k_{\perp}})$} & 
\colhead{$P_{shear}(k_{\parallel})$} &
\colhead{$P_{shear}({k_{\perp}})$} & 
\colhead{$P_{comp}(k_{\parallel})$} &
\colhead{$P_{comp}({k_{\perp}})$} & 
}
\startdata
A  &0.01     &4.5  &2.7  &5.3  &3.5  &5.5  &3.5  &5.5  &4.1 \\
B  &0.10     &4.0  &3.6  &4.3  &4.3  &4.3  &4.3  &4.3  &4.3 \\
C  &1.00     &3.6  &3.6  &4.9  &4.7  &5.0  &4.8  &4.5  &4.5 \\
D  &$\infty$ &N/A  &N/A  &4.8  &4.8  &4.9  &4.9  &4.7  &4.7 \\
A2 &0.01     &4.5  &3.0  &4.7  &3.4  &4.8  &3.3  &4.8  &4.0 \\
C2 &1.00     &3.3  &3.3  &4.3  &4.3  &4.4  &4.4  &4.2  &4.2 \\
 \enddata
\tablecomments{Values of spectral indices $n$ of power-law fits
   $\propto k^{-n}$ to $P(k_{\parallel})$ and $P(k_{\perp})$.}
\end{deluxetable}


\begin{thebibliography}{}

\bibitem[Biskamp \& M\"{u}ller(2000)]{Biskamp2000} Biskamp, D. \& M\"{u}ller, W.-C. 2000, Phys. Plasmas, 7, 4889

\bibitem[Blitz(1993)]{Blitz1993} Blitz, L. 1993, in Protostars and Planets III, ed. E.H. Levy \& J.I. Lunine (Tuscon: Univ. Arizona Press), pp. 125-161

\bibitem[Boldyrev(2002)]{Boldyrev2002a} Boldyrev, S. 2002, \apj, 569, 841

\bibitem[Boldyrev et al.(2002)]{Boldyrev2002b} Boldyrev, S., Nordlund, A., \& Padoan, P. 2002, \apj, 573, 678 

\bibitem[Brunt \& Heyer(2002)]{Brunt2002} Brunt, C.M. and Heyer, H.H. 2002, \apj, 566, 289 

\bibitem[Burgers(1974)]{Burgers74} Burgers, J.M. 1974, The Nonlinear Diffusion Equation (Dordrecht: Reidel)


\bibitem[Cho \& Lazarian(2002)]{Cho2002} Cho, J. \& Lazarian, A. 2002, Phys. Rev. Lett., 88, 245001

\bibitem[Cho et al.(2002a)]{Cho2002a} Cho, J., Lazarian, A. \& Vishniac, E.T. 2002a, \apj, 564, 291

\bibitem[Cho et al.(2002b)]{Cho2002b} Cho, J., Lazarian, A. \& Vishniac, E.T. 2002b, accepted for publication in ``Simulations of Magnetohydrodynamic Turbulence in Astrophysics'', Eds. T. Passot \& E. Falgarone, Springer Lecture Notes in Physics, astro-ph/0205286

\bibitem[Cho \& Vishniac(2000)]{Cho2000} Cho, J., \& Vishniac, E.T. 2000, \apj, 539, 273

\bibitem[Crutcher(1999)]{Crutcher99} Crutcher, R.M. 1999, \apj, 520, 706

\bibitem[Dubrulle(1994)]{Dubrulle94} Dubrulle, B., 1994, Phys. Rev. Lett., 73, 959

\bibitem[Goldreich \& Sridhar(1995)]{GS95} Goldreich, P., \& Sridhar, H. 1995, \apj, 438, 763 (GS)

\bibitem[Goldreich \& Sridhar(1997)]{GS97} Goldreich, P., \& Sridhar, H. 1997, \apj, 485, 680

\bibitem[Heyer \& Schloerb(1997)]{HS97} Heyer, M.H., \& Schloerb, F.P. 1997, \apj, 475, 173

\bibitem[Iroshnikov(1963)]{IK63} Iroshnikov, P. 1963, Soviet Astron., 7, 566

\bibitem[Kolmogorov(1941)]{K41} Kolmogorov, A. 1941, Dokl. Akad. Nauk SSSR, 31, 538

\bibitem[Kraichnan(1965)]{IK65} Kraichnan, R. 1965, Phys. Fluids, 8, 1385 

\bibitem[Larson(1981)]{Larson81} Larson, R.B. 1981, MNRAS, 194, 809

\bibitem[Lazarian \& Pogosyan(2000)]{LP2000} Lazarian, A., \& Pogosyan, D. 2000, \apj, 537, 720

\bibitem[Mac Low et al.(1998)]{MacLow1998} Mac Low, M.-M., Klessen, R.S., \& Burkert, A. 1998, Phys. Rev. Lett., 80, 2754

\bibitem[Mac Low \& Ossenkopf(2000)]{MacLow2000} Mac Low, M.-M., Ossenkopf, V. 2000, Astron. Astrophys., 353, 339

\bibitem[Maron \& Goldreich(2001)]{MG2001} Maron, J., \& Goldreich, P. 2001, \apj, 554, 1175

\bibitem[Matthaeus et al.(1996)]{MGOR96} Matthaeus, W.H., Ghosh, S., Oughton, S., \& Roberts, A.D. 1996, J. Geophys. Res., 101, 7619

\bibitem[Matthaeus et al.(1998)]{MOGH98} Matthaeus, W.H., Oughton, S., Ghosh, S. \& Hossain, M. 1998, Phys. Rev. Lett., 81, 2056

\bibitem[Milano et al.(2001)]{Milano2001} Milano, L.J., Matthaeus, W.H., Dmitruk,P. \& Montgomery, D.C. 2001, Phys. Plasmas., 8, 2673

\bibitem[M\"{u}ller \& Biskamp(2000)]{Muller2000} M\"{u}ller, W.-C., \& Biskamp, D. 2000, Phys. Rev. Lett., 84, 475

\bibitem[Ossenkopf \& Mac Low(2002)]{OM2002} Ossenkopf, V., \& Mac Low, M.-M. 2002, A \& A, 390, 307 

\bibitem[Ostriker et al.(1999)]{Ostriker1999} Ostriker, E.C., Gammie, C.F., \& Stone, J.M. 1999, \apj, 513, 259

\bibitem[Ostriker et al.(2001)]{Ostriker2001} Ostriker, E.C., Stone, J.M. \& Gammie, C.F. 2001, \apj, 546, 980

\bibitem[Ostriker(2002)]{Ostriker2002} Ostriker, E.C. 2002, accepted for publication in ``Simulations of Magnetohydrodynamic Turbulence in Astrophysics'', Eds. T. Passot \& E. Falgarone, Springer Lecture Notes in Physics, astro-ph/0204463

\bibitem[Oughton et al.(1994)]{OPM94} Oughton, S., Priest, E.R. \& Matthaeus, W.H. 1994, J. Fluid Mech., 280, 95

\bibitem[Oughton et al.(1998)]{OMG98} Oughton, S., Matthaeus, W.H. \& Ghosh, S. 1998, Phys. Plasmas, 5, 4235

\bibitem[Padoan \& Nordlund(1999)]{Padoan1999} Padoan, P., \& Nordlund, A. 1999, \apj, 526, 279

\bibitem[Porter et al.(1994)]{PPW94} Porter, D.H., Pouquet, A., \& Woodward, P.R. 1994, Phys. Fluids, 6, 2133

\bibitem[Porter et al.(1998)]{PWP98} Porter, D.H., Woodward, P.R. \& Pouquet, A. 1998, Phys. Fluids, 10, 237

\bibitem[Porter et al.(1999)]{PPSW99} Porter, D.H., Pouquet, A., Sytine, I. \& Woodward, P.R. 1999, Physica A 263 (1999), 263-270

\bibitem[Rosolowsky et al.(1999)]{Rosolowsky99} Rosolowsky, E.W., Goodman, A.A., Wilner, D.J., \& Williams, J.P. 1999, \apj, 524, 887

\bibitem[She \& L\'{e}v\^{e}que(1994)]{SL94} She, Z.-S., \& L\'{e}v\^{e}que, E.
1994, Phys. Rev. Lett., 72, 336

\bibitem[Shebalin et al.(1983)]{SMM83} Shebalin, J.V., Matthaeus, W.H., \& Montgomery, D. 1983, J. Plasma Phys., 29, 525

\bibitem[Solomon et al.(1987)]{Solomon87} Solomon, P.M., Rivolo, A.R., Barrett, J., \& Yahil, A. 1987, \apj, 319, 730

\bibitem[Stone \& Norman(1992a)]{Stone1992a} Stone, J.M., \& Norman, M.L.
1992a,  ApJS, 80, 753

\bibitem[Stone \& Norman(1992b)]{Stone1992b} Stone, J.M., \& Norman, M.L.
1992b, ApJS, 80, 791

\bibitem[Stone et al.(1998)]{Stone1998} Stone, J.M., Ostriker, E.C., \& Gammie, C.F. 1998, \apj, 508, L99 (Paper II)

\bibitem[V\'{a}zquez-Semandeni et al.(2000)]{Vazquez2000} V\'{a}zquez-Semandeni, E.,
 Ostriker, E.C., Passot, T., Gammie, C.F., \& Stone, J.M. 2000, in Protostars and Planets IV, ed. Mannings, Boss, \& Russell (Tuscon: Univ. Arizona Press)

\end{thebibliography}
\end{document}